\documentstyle[12pt,a4,epsfig]{article}

\newcommand{\be}{\begin{equation}}
\newcommand{\ee}{\end{equation}}
\newcommand{\ba}{\begin{eqnarray}}
\newcommand{\ea}{\end{eqnarray}}

\begin{document}

\begin{titlepage}
\begin{flushright}
{BUTP--97/19}\\
{hep--lat/9708005}\\
\end{flushright}
\vspace{1.7cm}
\begin{center}
{\large\bf Quenched Chiral Perturbation Theory to one loop}\\
\vspace{1.8cm}
{\bf Gilberto Colangelo$^{a,}$\footnote{e--mail: colangelo@lnf.infn.it}}
 and {\bf Elisabetta Pallante$^{b,}$\footnote{e--mail:
     pallante@butp.unibe.ch}  } 
\\[0.6cm]
$^{a}$ {\em INFN -- Laboratori Nazionali di Frascati,\\
P.O. Box 13, I--00044 Frascati (Rome), Italy}\\
\vspace{0.3cm}
$^{b}$ {\em University of Bern, Sidlerstrasse 5,\\
CH--3012 Bern, Switzerland}\\[0.5cm]
\end{center}
\vfill
\vspace{0.6cm}
\begin{abstract}
We calculate the divergences of the generating functional of quenched
Chiral Perturbation Theory at one loop, and renormalize the theory by
an appropriate definition of the counterterms. We show that the
quenched chiral logarithms can be accounted for by defining a renormalized
 $B_0$ parameter which, at lowest order, is
proportional to the vacuum expectation value of the scalar quark
density.
Finally, we calculate several quantities at one loop to better analyze the
modifications induced by quenching in the ultraviolet 
finite part of the one--loop
corrections. We point out that some of the finite loop corrections may
diverge in the chiral limit.
\end{abstract}
\vspace{1cm}
\hspace{0.6cm}{\small PACS: 12.39.Fe, 12.38.Gc.}
\vspace{0.5cm}

\vspace*{1cm}
\setcounter{equation}{0}
\setcounter{figure}{0}

\vfill
August 1997
\end{titlepage}

\renewcommand{\theequation}{\arabic{section}.\arabic{equation}}
\setcounter{equation}{0}
\section{Introduction}
Most lattice calculations of QCD in its non--perturbative regime and
weak interactions use at present the quenched approximation,
i.e. neglect the effect of virtual quark loops. Taking them into
account considerably increases computing times. 
This means that presumably the quenched approximation will remain with
us for quite a long time: even with computers much faster than those
presently available, it will always offer the chance to make a
low cost exploratory calculation before embarking on a full QCD
simulation. 

Simulations of quenched QCD would be much more useful if we had a real
understanding of the effects of this approximation.
Investigations in this direction have been made by several authors
\cite{Morel,Sharpe,qCHPT}. At present we see one main approach that has
proven to be the most systematic, and also to incorporate most of the
useful ideas that have been proposed on the subject. This method is
called quenched Chiral Perturbation Theory (qCHPT), and has been
originally proposed by Bernard and Golterman in Ref. \cite{qCHPT} for
the purely strong sector (strong interactions in the presence of
external fields). It has been recently extended to the heavy--light meson
sector \cite{HLmeson}, to vector mesons \cite{Vector} and to the baryon sector
 \cite{lasha}. It has been also used in the context of 
non--leptonic weak interactions \cite{Sharpe,Kpipi}.

Let us shortly review the main ideas behind this approach.
The difficulty to control the quenched
approximation comes from the fact that one is modifying the theory at
the non--perturbative level. On the other hand we know that at low
energy it is possible to define a perturbative scheme to study the
strong interactions: this scheme is known as Chiral Perturbation Theory 
(CHPT). In this framework the expansion parameter is given by the
energy of the weakly interacting Goldstone bosons of the spontaneously
broken chiral symmetry: these have a vanishing interaction at zero
energy, as symmetry dictates. The chiral symmetry imposes also a set 
of relations between the coefficients of this expansion in different
amplitudes. Those relations do not fully constrain the theory that at
each order of the expansion has a number of free constants. These
constants incorporate the effect of the non--perturbative QCD dynamics.
Under the assumption that in the quenched approximation the
mechanism of spontaneous chiral symmetry breaking is preserved, one
may attempt to construct a perturbative scheme for the quenched
case, analogous to the one valid in the full QCD case. 
In this manner one would be able to calculate those effects of
quenching that modify the perturbative, calculable part of the
theory. On the other hand, the changes in the unconstrained low energy
constants remain unknown, being due to the modifications which affect
the non--perturbative QCD dynamics.
This method has the advantage of introducing from the start this
clear, useful separation between the non--perturbative dynamics of the
fundamental theory and the perturbative, predictable dynamics of the
Goldstone bosons. 

A peculiar aspect of the quenched approximation 
comes  from the $U(1)$ axial anomaly of QCD. In the fundamental theory
the would--be--Goldstone boson (the $\eta^\prime$) does not become
massless in the chiral limit, since the axial anomaly generates a singlet
component (heavy) mass at the level of the effective theory. Thus, in
the real world the $\eta^\prime$ is heavy and
decoupled from the octet of the pseudo--Goldstone bosons. In the
quenched approximation this decoupling stops halfway: only one of the
diagrams that are responsible for the decoupling of the $\eta^\prime$
survives. At the level of the effective theory this has important
consequences:  
the singlet field remains light (degenerate with the Goldstone
bosons) and has to be treated on the same footing as the
octet fields. However its two--point function develops a double pole and
does not admit an interpretation as a propagator.
Treating the singlet as a dynamical degree of freedom brings
in new constants in the effective theory. One of them is a new mass
scale (the singlet mass $m_0$) that is generated by the anomaly, and
that does not vanish in the chiral limit. This mass appears in the
numerator of the double--pole term in the singlet two--point function.
As different authors have shown \cite{Sharpe,qCHPT}, this double pole  
is responsible for the presence of a new type
of chiral logarithms (we denote them as quenched chiral logs) in loop
corrections,  of the form $m_0^2 \ln M_\pi^2$, as opposed to the
standard ones $M_\pi^2 \ln M_\pi^2$.  
This is one of the main qualitative differences that arises from the
quenched version of CHPT.

So far, works in quenched CHPT have concentrated on specific processes,
analyzing the changes induced in Goldstone boson loops and the size of  
the effect of quenched chiral logarithms.
The aim of the present work is to perform a complete renormalization
of the theory at the one--loop level, on the same line as what has been
done by Gasser and Leutwyler in the ordinary CHPT case \cite{gl84,
  gl85}.  
This requires a calculation of all the ultraviolet
divergent pieces of the generating functional and a definition of the
Lagrangian at order $p^4$, the next--to--leading order.
The advantages of the present analysis are the following:
\begin{enumerate}
\item[1)]
The calculation of the divergences and renormalization can be
done for a generic number of flavours $N$. 
As we have shown in Ref. \cite{pl} the
$N$--dependence of the divergences can be used to verify the
cancellation of quark loops in the effective theory.
\item[2)]
Like in the standard case, the calculation of the divergences at the
generating functional level provides a useful check for single
amplitude calculations. This check is even more welcome in qCHPT where
the number of graphs to be computed becomes soon very large.
\item[3)]
This calculation allows to have full control on the divergences due to
singlet loops. In particular we will show that quenched
chiral logarithms can be accounted for via a renormalization 
of the low--energy constant $B_0$ (which is proportional to the
$\bar{q} q$ condensate). This constant appears in all other quantities
through the pion mass squared, with the only exception of $\bar{q}q$
matrix elements, that have it as an explicit factor.
\end{enumerate}
After having performed the one--loop renormalization, we will devote
our attention to the ultraviolet finite part of the one--loop corrections, by
computing specific physical quantities at one loop.
The relevance of the finite part of the loop corrections is in the
fact that they may contain terms which diverge in the chiral limit like an
inverse power of the quark mass. One can realize that this may happen
by simply looking at the standard chiral power counting
\cite{weinberg}, and taking into account the fact that in 
quenched CHPT a new vertex appears with chiral order zero (the vertex
proportional to $m_0^2$).
Power--like chiral divergences and quenched chiral logs are the crucial
problem of the quenched version of CHPT: the effective theory is
defined as an expansion around the chiral limit, and this  
limit is no more well defined in the quenched case. On the other hand
these divergences seem to be unavoidable in the present framework and
it looks plausible that they are a direct consequence 
of the sicknesses of quenched QCD. To clarify this very
important point, a direct evidence of these effects in lattice
simulations of quenched QCD would be most welcome.  

In our analysis of various observables we will give the complete one--loop
results. Our aim is not just to make predictions, or to compare with
numbers produced in lattice simulations.
Rather, we would like to show in detail how the quenched
approximation distorts the matrix elements. For this reason we
will only work in Minkowski space--time: all the formulae will be given
with the idea that one should be able to easily see the difference from the
corresponding ones calculated in standard CHPT.
In particular we will stress the presence of terms divergent in the
chiral limit and of unphysical threshold singularities in Minkowski
space--time at infinite volume. These type of singularities have been
already discussed in the literature
\cite{bg,pipiq}, and have lead to the conclusion that quenched CHPT makes
sense only in Euclidean space--time. Despite this, we still
prefer to calculate amplitudes in Minkowski space--time, considering them
as formal expressions. As we just said this will make the comparison to
standard CHPT amplitudes easier; on the other hand, the modifications
needed to go to Euclidean space--time can be easily implemented.

The plan of the paper is as follows. In Section 2 we outline the main
steps from CHPT to its quenched version. We give the leading order
Lagrangian and define our notation, both for CHPT and quenched
CHPT. In Section 3 we calculate the divergences of qCHPT to one loop
using the background field method, while Section 4 contains the list
of counterterms for a generic 
number of flavours $N$ and for $N=3$ and 2. This completes the
renormalization of the theory at the one--loop level.
In Section 5 we analyze a few quantities to one loop in the two
degenerate flavours 
case. These are the $\bar{q}q$ condensate, the pion mass and decay
constant, the scalar and vector form factors of the pion, and the $\pi
\pi$ scattering amplitude. 
In Section 6 we state our conclusions.
We have also three appendices. In Appendix A we give a simple
derivation of the divergent term proportional to $m_0^2$ in the
quenched generating functional. 
In Appendix B we give the explicit $N$--dependence of the divergences
in the non--leptonic weak interactions sector, and guess the divergences
in the quenched case by simply dropping any $N$--dependence.
Finally, in Appendix C we give the explicit expressions for the
one--loop functions which enter the calculations.

\renewcommand{\theequation}{\arabic{section}.\arabic{equation}}
\setcounter{equation}{0}
\section{From CHPT to its quenched version}
In this section we introduce the standard notation of Chiral Perturbation
Theory, that will be also used in its quenched version.
We work in Minkowski space--time in both cases for ease of comparison.
For any further detail in the derivation of the CHPT
Lagrangian we refer the reader to the original works by Gasser and
Leutwyler \cite{gl84,gl85}.
The construction of the CHPT Lagrangian is based on the identification
of the symmetry group of the QCD Lagrangian in the chiral limit,
which, for $N$ flavours is given by  $U(N)_L\otimes
U(N)_R$, and on the well supported assumption that the symmetry of the
subgroup $SU(N)_L\otimes SU(N)_R$ is spontaneously broken to $SU(N)_V$.
The extension of this construction to the quenched case was proposed
by Bernard and Golterman \cite{qCHPT} on the basis of an observation
made by Morel \cite{Morel}. He observed that, formally, a Lagrangian
corresponding to quenched QCD can be obtained by adding to the QCD
Lagrangian a term which is totally analogous to that for quark fields,
but which contains {\em ghost} spin--1/2 fields with wrong,
i.e. bosonic statistics. 
The symmetry of the resulting Lagrangian in the chiral limit is larger
than that of QCD and is given by the graded group: $U(N|N)_L\otimes
U(N|N)_R$, describing transformations between $N$ physical
flavours and $N$ ghost flavours. 
It is then assumed that also in the quenched case a
spontaneous symmetry breaking down to the diagonal subgroup
$SU(N|N)_V$ occurs. Like in standard QCD, the $U(1)_A$ symmetry is
anomalous. 

\subsection{Standard CHPT}
Chiral Perturbation Theory describes the dynamics of the octet Goldstone
bosons fields (pions) of the spontaneously broken chiral symmetry of
QCD. It is an expansion in powers of the energy of the Goldstone
bosons and the light quark masses. 
The lowest order CHPT Lagrangian, i.e. at order $p^2$ and linear in
the quark masses, can be written in the following form:
\be
{\cal L}_2 = \frac{F^2}{4} \langle D_\mu U D^\mu U^\dagger + U^\dagger 
\chi + \chi^\dagger U \rangle  
= \frac{F^2}{4} \langle u_\mu u^\mu + \chi_+ \rangle \; ,
\label{LCHPT}
\ee
where $\langle\ldots\rangle$ stands for the trace over flavour
indices, $F$ is the bare pion decay constant and the fields are
defined as follows  
\ba
U= u^2 &=& \exp\left(\sqrt{2} i \phi/F \right) \; , \nonumber \\
D_\mu U &=& \partial_\mu U -i r_\mu U + i U l_\mu \; , \nonumber \\
\chi &=& 2 B_0(s+ip) \; ,\nonumber\\
u_\mu&=&iu^\dagger D_\mu U u^\dagger = u_\mu^\dagger \; , \nonumber \\
\chi_+ &=& u \chi^\dagger u + u^\dagger \chi u^\dagger \; .
\label{DEFF}
\ea
The Lagrangian contains the external sources $s, p, v_\mu, a_\mu$,
$r_\mu=v_\mu+a_\mu$, $l_\mu=v_\mu-a_\mu$, which are $N\times N$
matrices, with $N$ the number of flavours.
The field $\phi$ is an $N\times N$ matrix that contains the Goldstone
bosons fields: $\phi =1/\sqrt{2} \sum_{i=1}^{N^2-1} \lambda_i
\phi^i$. In case, one may add to $\phi$ a 
singlet component, so that $\langle \phi \rangle = \phi_0$.
In the presence of a singlet component the Lagrangian in
Eq. (\ref{LCHPT}) is invariant under $U(N)_L\otimes U(N)_R$. Since in
QCD the  $U(1)_A$ subgroup is anomalous the breaking pattern
$U(N)_L\otimes U(N)_R\to SU(N)_L\otimes SU(N)_R\otimes U(1)_V$ is
realized. The invariance under the residual unbroken group allows for
the presence of extra functions of the singlet component $\phi_0$
only. A possible choice for this Lagrangian, compatible with $P,C,T$
and chiral invariance, is (see also \cite{gl85} for a different
choice): 
\be
{\cal L}_2= V_1(\phi_0) \langle D_\mu U D^\mu U^\dagger\rangle +V_2(\phi_0) 
\langle  U^\dagger \chi + \chi^\dagger U\rangle -V_0(\phi_0) 
+V_5(\phi_0) D_\mu\phi_0 D^\mu \phi_0  \; ,
\label{SCHPT}
\ee
where all the functions $V_i$ are even and real functions of $\phi_0$.

\subsection{Quenched CHPT}
The modification needed to construct the quenched version of the CHPT 
Lagrangian in Eq. (\ref{SCHPT}) for a generic number of flavours $N$ 
consists of the extension of the chiral symmetry group 
$SU(N)_L\otimes SU(N)_R\otimes U(1)_V$ to the graded group
$\left[ SU(N|N)_L\otimes SU(N|N)_R \right] \odot U(1)_V$, which enlarges
the spectrum of the theory to include {\em ghost} states 
(the $\odot$ stands for the semidirect product of $U(1)_V$,
 which does not commute with transformations that exchange
particles with ghosts). In the quenched case there are $N$ physical
flavours and $N$ ghost flavours. Under the graded extension all the
$N\times N$ matrices representative of the original $U(N)$ group are
transformed into graded $2\times 2$ block matrices  
\[
A \rightarrow \left( 
\begin{array}{cc}
A & B  \\
C & D
\end{array}  \right) \; ,
\]
whose components are in turn $N \times N$ matrices.
The matrices $A$ and $D$ ($B$ and $C$) have bosonic (fermionic) character.
The trace is then transformed into supertrace:
\[
\mbox{tr} (A) \rightarrow \mbox{str} \left( 
\begin{array}{cc}
A & B  \\
C & D
\end{array}  \right) = \mbox{tr}(A)-\mbox{tr}(D) \; . 
\]
The leading order Lagrangian of quenched CHPT can be written in full analogy 
to the standard CHPT case\footnote{To distinguish between a quenched CHPT
  quantity and its standard counterpart we use either capital letters
  (as in $\phi \rightarrow \Phi$) or, when this is not possible, the
  $s$ subscript (as in $U \rightarrow U_s$).}:
\ba
{\cal L}_2&=&V_1(\Phi_0) \mbox{str} (D_\mu U_sD^\mu U_s^\dagger )+V_2(\Phi_0) 
\mbox{str} (\chi^\dagger_s U_s+U_s^\dagger\chi_s )
-V_0(\Phi_0) \nonumber\\
&&+V_5(\Phi_0) D_\mu\Phi_0 D^\mu \Phi_0 \;  ,
\label{L2}
\ea
where again $V_i(\Phi_0)$ are even and real functions of the generalized
singlet field $\Phi_0$.
The graded meson field is defined through the usual exponential
representation: 
\[
U_s = \exp (\sqrt{2} i\, \Phi/F) \; \; ,
\]
where $F$ is the bare quenched pion decay constant and $\Phi$ is now a 
hermitian non traceless $2\times 2$ block matrix
\[
\Phi = \left( 
\begin{array}{cc}
\phi & \theta^\dagger  \\
\theta & \tilde\phi
\end{array}  \right) \; , ~~~\mbox{str}(\Phi) = \Phi_0 =
\phi_0-\tilde\phi_0 \; \; ,
\]
which contains the new {\em ghost} states of the quenched
spectrum. All the possible quenched meson states carry the quantum
numbers of a two particle bound state made up with quarks $q$ or
ghost--quarks $\tilde{q}$. On the diagonal sites it contains the
physical pseudo--Goldstone boson matrix $\phi$ (i.e. the physical pions
including the singlet component), with the quantum numbers of a
$q\bar{q}$ pair, and the ghost field matrix $\tilde{\phi}$, with the
quantum numbers of a $\tilde{q}\bar{\tilde{q}}$ pair. They are both of
bosonic nature. In the off--diagonal sites are the {\em ghost} hybrid
fields $\theta$ and $\theta^\dagger$, which carry the quantum numbers of a
mixed $\tilde{q}\bar{q}$ and $q\bar{\tilde{q}}$ pair respectively, 
both of fermionic nature. 
This spectrum of meson states can be found also in the
original derivation by Morel \cite{Morel}. He calculated the
functional integral over the quark and ghost--quark fields (in the
leading large--$d$ expansion and strong gauge coupling limit) and
obtained exactly the meson spectrum of quenched CHPT, with
mesons of composite nature, given by bilinears of quarks/ghost--quarks
at the same lattice site.

The covariant derivative over the field $U_s$ is defined as 
$D^\mu U_s = \partial^\mu U_s - i r_s^\mu U_s+iU_s l_s^\mu$, where $r_s^\mu
(l_s^\mu )$ is the right(left)--handed external source of the graded group.
The field $\chi_s=2B_0(s_s+ip_s)$ contains the external scalar ($s_s$)
and pseudoscalar ($p_s$) sources analogously to the ordinary CHPT
case. All the external fields $r_s^\mu , l_s^\mu , s_s, p_s$ are
generalizations of the standard external fields, in order to make the
Lagrangian in Eq. (\ref{L2}) locally invariant under the graded group
$\left[ SU_L(N|N)\otimes SU_R(N|N) \right] \odot U(1)_V$. 
Since we are not interested in studying matrix elements containing the
spurious fields as external legs, we will always use the standard
external sources only. With this reduction a generic graded source
reads as follows:
\[
j_s = 
\left(
\begin{array}{cc} 
j  & 0 \\
0 & 0
\end{array} 
\right) \; ,\; j= p,v_\mu, a_\mu \; \; .  
\]
For the scalar external source we must recall that it is defined to
contain the quark mass matrix ${\cal M }$ which has to be the same
both for the quarks and the ghosts:
\[  
s_s=\left(  
\begin{array}{cc} 
{\cal M } + \delta s  & 0  \\
0 & {\cal M} 
\end{array}  \right)  \; \; .
\] 
In what follows the quark mass matrix will be taken
proportional to the unit matrix: ${\cal M} = m_q { \bf 1}$. All the
Goldstone bosons and their ghost counterparts will have the same 
mass: $M^2=2 B_0  m_q$. We have adopted the usual CHPT notation and
call $M^2$ the lowest order term in the expansion of the mass of the
pions in powers of quark masses:
\be
M_\pi^2 = 2 B_0 m_q + O(m_q^2) = M^2 + O(m_q^2) \; \; .
\ee
Finally, we expand the functions $V_i(\Phi_0)$ in powers of $\Phi_0$:
\ba
V_0(\Phi_0) &=& \frac{m_0^2}{2N_c} \Phi_0^2 +O(\Phi_0^4) \; \; ,
\nonumber \\ 
V_{1,2}(\Phi_0) &=& \frac{F^2}{4}+{1\over 2}v_{1,2}\,\Phi_0^2 +
O(\Phi_0^4) \; \; , \nonumber \\
V_5(\Phi_0) &=& \frac{\alpha}{2N_c} + O(\Phi_0^2) \; \; ,
\ea
and we shall always work with number of colours $N_c=3$.

\renewcommand{\theequation}{\arabic{section}.\arabic{equation}}
\setcounter{equation}{0}
\section{One--loop divergences}

To calculate the ultraviolet divergent part of the quenched generating 
functional to one
loop we use the background field method, i.e. expand the action around the
classical solution, which is determined by the external sources through the 
classical equations of motion.
We write the field $U_s$ as:
\[
U_s= u_s ~e^{i \Xi}~ u_s \; \; ,
\]
where $\bar{U_s}=u_s^2 $ is the classical solution to the
equations of motion. In the absence of spurious external sources it
reduces to 
\[  
u_s=\left(  
\begin{array}{cc} 
{ u }   & 0  \\
0 & {\bf 1} 
\end{array}  \right)  \; \; .
\]
We decompose the fluctuation $\Xi$ similarly to the field $\Phi$ and write:
\[
\Xi = \left( 
\begin{array}{cc}
\xi & \zeta^\dagger  \\
\zeta & \tilde\xi
\end{array}  \right) \;\; , \; \; \; \; \mbox{str}(\Xi)=
\sqrt{N}(\xi_0-\tilde\xi_0) \; \; ,
\]
(note that with this normalization the $\xi_0$ and $\tilde\xi_0$ have
a proper kinetic term).
The matrix fields $\xi$ and $\zeta$
are decomposed as follows
\be
\xi =\sum_{a=0}^{N^2-1} \hat\lambda_a \xi^a , 
\; \; \zeta =
\sum_{a=0}^{N^2-1} \hat\lambda_a \zeta^a  \; , 
\ee
and the fields $\tilde\xi$ and $\zeta^\dagger$ analogously,
where $\hat\lambda_a = \lambda_a/\sqrt{2}, \; \; a=1,\ldots,N^2-1$, and
$\hat\lambda_0= {\bf 1}/\sqrt{N}$.
Given their special character, it is useful to separate the singlet
components of the $\xi$ and $\tilde\xi$ fields from the rest, and
combine them into one vector:
\[
X_0 = \left( 
\begin{array}{c}
\xi_0 \\
\tilde\xi_0
\end{array}  \right) \; \;.
\]
The remaining fields are put into components of the following
vectors: 
\[
\xi^T = (\xi^1, \; \xi^2, \ldots, \; \xi^{N^2-1})\;, ~~~~ \zeta^\dagger =
(\zeta^{\dagger \;0}, \; \zeta^{\dagger \;1}, \ldots, \;
\zeta^{\dagger \; N^2-1}) \; .
\]
With this notation the action can be written as
\ba
S[\Phi] &=& S[\bar{\Phi}] - \frac{F^2}{4} \int dx \left\{ 
X_0^T D_X X_0 +
  \xi^T D_\xi \xi + \xi_0 B^T \xi + \xi^T B \xi_0 + 
2 \zeta^\dagger D_\zeta \zeta
\right. \nonumber \\ 
&-& \left. \tilde{\xi}^T (\Box + M^2) \tilde{\xi}\right\} +
O(\Xi^3) \; \; .
\ea
The explicit expressions for the various differential operators
$D_{X,\xi,\zeta}$ and the matrix $B$ will be given below.
The matrix $B$ induces a mixing between the singlet and non singlet
component of the physical meson field. 
Notice also that the fields $\tilde\xi$ are completely decoupled from
the rest: 
the integration over these degrees of freedom produces only an
irrelevant constant.

Before deriving the various contributions to the generating
functional at one loop we shift the field $\xi$ in order to
remove the mixing with the singlet component $\xi_0$. By performing the
translation 
\[
\xi = \xi'- D_\xi^{-1} B \xi_0\; ,
\]
one gets
\ba
\xi^T D_\xi \xi + \xi_0 B^T \xi+ \xi^T B \xi_0 = 
\xi'^T D_\xi \xi'- \xi_0 B^T D_\xi^{-1} B
\xi_0\; .
\ea
In this manner the action up to the quadratic fluctuations becomes a sum of
quadratic differential forms diagonal in the fields $X_0, \; \xi, \;
\tilde\xi , \; \zeta, \; \zeta^\dagger$. The price to pay is that now
the differential operator acting on the singlet field $X_0$ has a
nonlocal term.  
Denoting as $\bar{D}_X$ the new non local operator acting on
$X_0$ after the shift, the quenched generating functional to one loop can be
formally written as follows
\be
e^{iZ^{\mbox{\tiny{qCHPT}}}_{\mbox{\tiny{1~loop}}}} = 
{\cal N} {\det D_\zeta\over (\det D_\xi )^{1\over 2}(\det
  \bar{D}_X)^{1\over 2}} \; \;.
\label{ZQ}
\ee
As we will see below, the non locality of $ \bar{D}_X$ will hardly make the 
calculation of the divergent part more complicated.

\subsection{Integral over the $\xi$ fields}

The differential operator $D_\xi^{ab}$ is defined as
follows\footnote{We remind the reader that the indices $a,\;b$ run
  from 1 to $N^2-1$. The singlet components are treated separately.}
\begin{eqnarray}
D_\xi^{ab}\xi_b &=& d_\mu d^\mu\xi^a+\hat{\sigma}^{ab}\xi_b \; \; ,
\nonumber\\ 
d_\mu\xi^a &=& \partial_\mu\xi^a+\hat{\Gamma}_\mu^{ab}\xi_b \; \; ,
\label{D}
\end{eqnarray}
where
\begin{equation}
\hat{\Gamma}_\mu^{ab}=-\langle \Gamma_\mu
[\hat\lambda^a,\hat\lambda^b ] \rangle \; \;, \; \; \; \;
\hat{\sigma}^{ab}= -{1\over 4} \langle [u_\mu ,\hat\lambda^a ] 
[u^\mu ,\hat\lambda^b ] \rangle +{1\over 4} \langle
\{\hat\lambda^a,\hat\lambda^b\}\chi_+ \rangle \; \; ,
\end{equation}
and $\Gamma_{\mu }= 1/2 ( [u^\dagger ,\partial_\mu u ]-i u^\dagger
r_\mu u -i u l_\mu u^\dagger )$ is the vector current connection of the
covariant derivative over the dynamical fields.

The ultraviolet divergent part of the integral over the $\xi$ fields can be
derived in closed form by regularizing the determinant in $d$
dimensions and using standard heat--kernel techniques. 
The result reads:
\begin{eqnarray}
{i \over 2} \ln \det D_\xi &=& -\frac{1}{(4\pi)^2(d\!-\!4)} \int \! dx
\left\{ {N \over 6} 
\langle \Gamma_{\mu \nu} \Gamma^{\mu \nu} \rangle + {1 \over 2}\left[
{1 \over 4} \langle u_\mu u_\nu \rangle \langle u^\mu u^\nu \rangle +
{1 \over 8} \langle u_\mu u^\mu \rangle^2 \right. \right. \nonumber \\
 &+&  {N \over 8} \langle (u_\mu u^\mu)^2 \rangle  + {N \over 4}
\langle  u_\mu u^\mu \chi_+ \rangle +  
{1 \over 4} \langle  u_\mu u^\mu \rangle \langle \chi_+ \rangle 
\nonumber\\
 &+& 
\left({N \over 8}-{1 \over {2N}} \right) \langle \chi_+^2 \rangle + 
\left({1 \over 8}+{1 \over {4N^2}} \right) \langle \chi_+ \rangle^2
 \nonumber\\ 
 &-& \left.\left. {1\over 2}\langle  u_\mu \rangle\langle u^\mu \left(
u_\nu u^\nu + 
\chi_+ \right) \rangle \right] \right\} +\ldots \; \; ,
\label{lndet1}
\end{eqnarray}
where the ellipsis stands for contributions which are finite in four
dimensions. This result is the standard CHPT result derived in 
 \cite{gl85}, where now we also keep terms proportional to $\langle
u_\mu\rangle$ that are nonzero only in the presence of the singlet
component. 

\subsection{Integral over the $\zeta$ fields}

The differential operator $D_\zeta^{ab}$ 
is defined like in Eq. (\ref{D}), but
with barred quantities, given by\footnote{Here the singlet component
  is included, and the indices $a,\;b$ run from 0 to $N^2-1$.} 
\begin{equation}
\bar{\Gamma}_\mu^{ab} =  -\langle\Gamma_\mu \hat\lambda^a \hat\lambda^b
\rangle \; \;, \; \; \; \;
\bar{\sigma}^{ab}=  {1\over 4} \langle( u_\mu u^\mu +\chi_+ +4B_0{\cal
M}) \hat\lambda^a
\hat\lambda^b \rangle  \; \; ,
\end{equation}
where we recall that ${\cal M}$ is the quark mass matrix.

The ultraviolet divergent part of the functional integral over 
the $\zeta$ fields
can also be given in closed form using standard heat--kernel
techniques. The result reads:
\be
i\ln \det D_\zeta=\frac{-1}{(4\pi)^2(d\!-\!4)} \! \int \! dx 
\left[ {N \over 6}
\langle \Gamma_{\mu \nu} \Gamma^{\mu \nu} \rangle +{N \over 16}
\langle (u_\mu u^\mu + \chi_+ + 4 B_0 {\cal M} )^2 \rangle \right]
+\ldots \; \; .
\label{lndetg}
\ee
As we remarked in Ref. \cite{pl}, the
integral over the $\zeta$ fields completely removes the terms linear
in $N$ in the divergences of standard CHPT to one loop. This
dependence is not fully explicit in Eq. (\ref{lndet1}), since a factor
$N$ is contained in the trace of $\chi_+$, when we expand this
around $s={\cal M}$ and for ${\cal M}$ diagonal:
\[
\langle \chi_+ \rangle = 2 N M^2 +O(\phi^2) \; \; .
\]
This result shows that the qCHPT scheme is coherent: the terms linear 
in  $N$ can only be generated by quark loops, and these are supposed
to be absent in the quenched approximation.

\subsection{Integral over the $X_0$ fields}

After the shift of the $\xi$ field the operator acting on $X_0$ can be
written as:
\be
X_0^T \overline{D}_X X_0 = X_0^T \left[ D_X-\frac{1}{2} (1+\tau_3) B^T
  D_\xi^{-1} B \right] X_0 \; \; ,
\ee
where
\ba
D_X &=& D_X^0+A_X \; \; ,\nonumber \\
D_X^0&=& \tau_3(\Box +M^2)+\frac{N}{3}(1-\tau_1)(\alpha \Box +m_0^2)
\; \; , \nonumber \\
A_X &=& \frac{1}{4N}(1+\tau_3) \langle \hat\chi_+ \rangle
 -N(1-\tau_1)\left ( v_1 \langle u_\mu u^\mu \rangle +v_2 \langle
\hat\chi_+ \rangle \right ) +O(\Phi_0^2) \; \; , \nonumber \\
B^a &=& \frac{1}{2\sqrt{2N}} \langle \lambda^a \chi_+ \rangle \; \; ,  
\label{DX}
\ea
and $\hat\chi_+=\chi_+-2M^2\bf{1}$, so that $\langle\hat\chi_+\rangle
=\langle\chi_+ \rangle -2NM^2$.
The expression of $D_X^0$, the ``free'' part of the differential
operator, clearly shows that the theory has a problem here: it is not
possible to diagonalize that operator, and we do not have two
freely propagating normal fields $(\xi_0, \tilde\xi_0 )$. 
On the other hand this problem is
welcome in this context, since it is thought to be the manifestation
of the absence of quark loops in the singlet field propagator, at the
level of the effective theory. In the language of Feynman diagrams
this problem shows up as a double pole in the propagator of the
singlet field, whose consequences on observables have been studied by
several authors \cite{Sharpe,qCHPT,bg}.
We adopt the usual point of view on this problem, i.e. assume that it
has to be there, and proceed with the calculation of the divergent
part of the generating functional.

In this case we cannot apply straightforwardly the heat--kernel
techniques, because the differential operator does not reduce to a
diagonal Klein--Gordon operator when the external fields are put to
zero. Therefore we just expand the logarithm of the differential
operator, and isolate the ultraviolet divergent terms:
\ba
\mbox{Tr} \ln \left(\overline{D}_X/D_X^0 \right)  &=& 
\mbox{Tr}\left[ {D_X^0}^{-1} (\overline{D}_X\!-\!D_X^0) \right]
\nonumber \\ 
&-&\frac{1}{2}
\mbox{Tr}\left[ {D_X^0}^{-1} (\overline{D}_X\!-\!D_X^0){D_X^0}^{-1}
  (\overline{D}_X\!-\!D_X^0) \right] + \ldots  \; .
\label{logDX}
\ea
One can easily see that the ellipsis in (\ref{logDX}) contains ultraviolet 
finite terms only. We postpone a more detailed discussion of the 
infrared behaviour of Eq. (\ref{logDX}) to the end of this section. 
The inverse of the ``free'' operator $D_X^0$ is:
\be
{D_X^0}^{-1} = G_0 \left[ \tau_3 - (1+\tau_1) \frac{N}{3} ( \alpha
  \Box + m_0^2 ) 
  G_0 \right] \; \; ,
\label{DX0-1}
\ee
where
\be
( \Box +M^2 )_x G_0(x-y) = \delta(x-y) \; \; ,
\ee
and 
\be
\overline{D}_X-D_X^0 = A_X-\frac{1}{2} (1+\tau_3) B^T D_\xi^{-1} B \;
\; . 
\label{ABX}
\ee
As we anticipated above, the overall effect of the shift made to
remove the mixing between singlet and nonsinglet fields is easily
accounted for. Expanding around the free part of $D_\xi^{-1}$ in the
non local term one gets 
\be
B^T D_\xi^{-1} B = \frac{1}{4N} G_0 \left[ \langle \chi_+^2 \rangle
  -\frac{1}{N} \langle \chi_+ \rangle^2 \right] +O(G_0^2) \; \; .
\ee
The term proportional to $O(G_0^2)$ can only yield ultraviolet 
finite contributions
to Eq. (\ref{logDX}), while the $G_0$ term yields ultraviolet divergent
contributions only to the first term of the expansion in
Eq. (\ref{logDX}). This shows that also 
in the singlet sector the UV divergent part is local and can be given in
closed form. 
The calculation of the ultraviolet divergent part of $\ln
\det \overline{D}_X$ is now easy: we simply have to insert back
Eqs. (\ref{DX0-1},\ref{ABX}) into Eq. (\ref{logDX}) and keep only the
UV divergent parts. Having worked out the traces (over the $\tau$
matrices) we obtain:
\ba
\frac{i}{2} \mbox{Tr} \ln \left( \overline{D}_X/D_X^0 \right) &=&
-\frac{1}{(4\pi)^2 (d-4)} \left\{ \frac{1}{4N} \langle \chi_+^2 \rangle
  -\frac{1}{8N^2} \langle \chi_+ \rangle^2 \right. \nonumber \\
&&  + \frac{m_0^2}{6} \langle
  \chi_+ \rangle - \frac{\alpha}{12} \langle \chi_+^2 \rangle +
  \frac{\alpha^2}{72} \langle \hat\chi_+ \rangle^2 \nonumber \\ 
&& -\left. \frac{1}{2} \langle \hat\chi_+ \rangle 
  \left(v_1 \langle u_\mu u^\mu \rangle +v_2 \langle
\hat\chi_+ \rangle \right) \right\} + \ldots \; \; ,
\label{Zsinglet}
\ea
where the ellipsis contains UV finite terms only.
The terms proportional to inverse powers of $N$ exactly cancel those
contained in Eq. (\ref{lndet1}) giving a result that is totally
$N$--independent. The terms proportional to $m_0^2$ and powers of
$\alpha$ are the effect of the double pole in the singlet propagator,
and are also $N$--independent. 
Note that no mixed terms of the type $(m_0^2,\alpha )\times (v_1,
v_2)$ can be produced in the divergent part.  

The term proportional to $m_0^2$ is a term already present in the
$O(p^2)$ Lagrangian. To remove that divergence one has to add to the
lowest order parameter $B_0$ in the ${\cal L}_2$ Lagrangian a
$d$--dependent part proportional to $m_0^2$ that has a pole at $d=4$: 
\be
B_0 \rightarrow B_0\left[1 + {\mu^{d-4}\over 16 \pi^2} \frac{1}{d-4}
  \frac{2m_0^2}{3F^2}+b_0(\mu)  \right]\; .
\label{B0ren}
\ee
This feature is completely new with respect to standard CHPT (in
dimensional regularization), and stems from the fact that in the quenched
theory we have a new mass scale that does not vanish in the chiral limit.
After the divergence has been removed, we are left with a term of the
form $m_0^2 \ln M^2 \langle \chi_+ \rangle$. This term contains all
the one--loop quenched chiral logs that have been discussed at length in
the literature. Our calculation shows that they can be fully
accounted for by defining a renormalized constant $\overline{B}_0$
\be
B_0 \rightarrow \overline{B}_0 = B_0 \left( 1- \frac{m_0^2}{48
   \pi^2 F^2}  \ln \frac{M^2}{\mu^2} +b_0(\mu)\right) \; \; .
\label{B0bar}
\ee
Notice that since $B_0$ is independent from the quark masses,
$\overline{B}_0$ becomes divergent in the chiral limit.
To find evidence for these quenched chiral logs one
should try to extract from lattice data this quantity
$\overline{B}_0$. As we will see the quark condensate and the
scalar form factor are two excellent candidates for this, since they are
the simplest quantities which are explicitly proportional to
$\overline{B}_0$. Other quantities will tipically depend on
$\overline{B}_0$ through the renormalized pion mass. This at one loop is
given by: 
\be
M_\pi^2 = 2 \overline{B}_0  m_q +O(m_q^2) \; \; ,
\ee
and is not divergent in the chiral limit. These other quantities are
therefore much less suitable to identify the presence of quenched chiral
logs. 

Of course what we have just said is valid in the specific sector we
are studying here. To extend it to other sectors of the effective
theory (like the non--leptonic weak interactions) requires further
study. However we have a rather simple argument that shows that what
has happened here will happen also in other sectors: the quenched
chiral logs to one loop contribute to the redefinition of one of the constants
appearing in the lowest order Lagrangian. In order not to interrupt
the discussion here we relegate the argument to Appendix \ref{m0}. 

\subsection{Complete result}
In this section we put together all the various pieces and
give the complete result for the ultraviolet divergent part of the generating
functional of qCHPT to one loop. The explicit expression for
Eq. (\ref{ZQ}) is: 
\ba
Z^{\mbox{\tiny{qCHPT}}}_{\mbox{\tiny{1~loop}}} &=& 
-\frac{1}{(4\pi)^2(d\!-\!4)} 
\int dx \left[
{1 \over 8} \langle u_\mu u_\nu \rangle \langle u^\mu u^\nu \rangle +
{1 \over 16} \langle u_\mu u^\mu \rangle^2 \right. \nonumber \\
&& + {1\over 8} \left(1 - 4 v_1 \right) \langle  u_\mu
u^\mu \rangle \langle \hat\chi_+ \rangle 
 + {1\over 16} \left(1 - 8 v_2 \right)\langle \hat\chi_+
 \rangle^2 \nonumber \\ 
&&+{m_0^2\over 6}\langle \chi_+ \rangle
 +{\alpha^2\over 72} \langle \hat\chi_+ \rangle^2
-{\alpha\over 12} \langle \chi_+^2 \rangle 
\nonumber\\
&& \left.- {1\over 4}\langle  u_\mu \rangle \langle u^\mu \left(u_\nu
    u^\nu + \chi_+\right) \rangle  
\right] +\ldots \; \; .
\label{ZZ}
\ea
The most striking feature of Eq. (\ref{ZZ}) is the complete flavour
independence of the result. 
If we analyze in detail the modifications that the quenched
approximation has produced to the divergent structure of the
effective theory at the one--loop level, we come to the following list:
\begin{enumerate}
\item
all the terms proportional to $N$ have been dropped;
\item
all the terms proportional to $1/N$ and $1/N^2$ have been dropped;
\item
new divergences proportional to the parameters present in the
anomalous singlet sector have been produced.
\end{enumerate}
All these new parameters are dimensionless, with the only exception of
$m_0$. The dimensionless parameters ($\alpha$ and $v_{1,2}$) generate
divergences that have the structure of a chiral invariant term (since
they do not break the chiral symmetry) of order $p^4$, for obvious
dimensional reasons. 
For the same reasons $m_0^2$ generates divergences with the structure of
a chiral invariant of order $p^2$. 
As it is shown in Appendix \ref{m0} one can very easily understand
why it is only the mass term $\langle \chi_+ \rangle$ that is
generated. 

As it turns out, the modifications listed in points 1. to 3. above 
find a very simple explanation: dropping the terms proportional to $N$ 
corresponds to
dropping virtual quark loops. Dropping the terms proportional to $1/N$ and
$1/N^2$, is a consequence of having a singlet degenerate in mass with
the nonsinglet pseudoscalars.
The new parameters in the singlet sector are required by the $U(1)_A$
anomaly, and the diseases in that sector are generated by the quenched
approximation, as it is well known.
These simple conclusions suggest that one could have guessed all these
modifications without doing any calculation.
In fact, we provide an example of how one could try such a guess in
Appendix \ref{NWI}, where we apply the same criteria to the
generating functional of the non--leptonic
weak interaction sector for the octet on--shell case (the complete
analysis will be given elsewhere \cite{weak}), by going through the
three steps we have enumerated above. 

\subsection{Chiral and threshold divergences}
Quenched chiral logs are not the only problem generated by the presence of
the double pole in the quenched version of the singlet propagator. 
As we will see in detail in Sec. 5 through several examples, this double pole
generates also other kind of divergences inside contributions which are
ultraviolet finite. These divergences are of two types: powerlike chiral
divergences, i.e. inverse powers of $M_\pi^2$, and unphysical threshold
divergences. 
We find it instructive, before closing this section to identify which are
the terms in the generating functional which are responsible for them.

Some of the terms (and in fact an infinite series of them) that we have
neglected in Eq. (\ref{Zsinglet}) because they are ultraviolet finite, 
contain these kind of singularities. They can be given in closed form only
if one stops at a given order in the expansion in powers of the field 
$\Phi$. Since 
in the following sections we are not going to analyze anything beyond the
four--point function, we can stop at order $\Phi^4$, and identify
explicitly the troublesome terms. They all come from the insertion of the
double pole term of (\ref{DX0-1}) in the expansion  (\ref{logDX}), and give
the following contribution to the generating functional:
\ba
\delta Z_{\mbox{\tiny{1 loop}}}^{\mbox{\tiny{qCHPT}}} 
&=& {(m_0^2-\alpha M^2)\over 24} \int \! dx
dy~\tilde{I}_1(x-y) 
\langle\hat{\chi}_+(y)\hat{\chi}_+(x)\rangle \nonumber\\
&-& \alpha {(m_0^2-\alpha M^2)\over 72} \int \! dx
dy~\tilde{I}_1(x-y) 
\langle\hat{\chi}_+(y)\rangle\langle\hat{\chi}_+(x)\rangle \nonumber\\
&-& {(m_0^2-\alpha M^2)^2\over 144}  \int \! dx dy~\tilde{I}_2(x-y)
\langle\hat{\chi}_+(y)\rangle\langle\hat{\chi}_+(x)\rangle\!+\!O(\!\Phi^6\!)
\; .  
\label{IRdiv}
\ea
The functions $\tilde{I}_1(z), \tilde{I}_2(z)$ are defined in appendix 
\ref{APP3}. At infinite volume and in Minkowski space--time their Fourier
transforms  ${I}_1(q^2), {I}_2(q^2)$ develop an imaginary part when $q^2
\geq 4 M_\pi^2 $ which diverges at $q^2 = 4 M_\pi^2$ (see appendix
\ref{APP3}). Moreover their values at $q^2=0$ are inversely proportional to
$M_\pi^2$ (again see appendix \ref{APP3}): this is the origin of 
powerlike chiral divergences that we will find in several observables in
Sect. 5. The threshold singularities in particular make the theory
meaningless in Minkowski space--time at infinite volume.
In finite volume and in Euclidean space--time the same one loop functions
${I}_{1,2}(q)$ have been evaluated at $q^2=4 M_\pi^2$ in
Ref. \cite{bg}, and it was found that these functions give rise to 
{\em enhanced finite volume} corrections which are forbidden in a healthy
Hamiltonian theory. As it was pointed out in Ref. \cite{bg} this shows that
qCHPT can only make sense in Euclidean space--time and in finite volume.

\renewcommand{\theequation}{\arabic{section}.\arabic{equation}}
\setcounter{equation}{0}
\section{Lagrangian at order $p^4$}
To complete the renormalization of the quenched theory at order $p^4$ one
needs to add the most general chiral invariant Lagrangian at this order. 
As in the
standard CHPT case, some of the couplings appearing in the order $p^4$ 
Lagrangian have an UV divergent part in such a way that all the one--loop 
divergences are removed.
The most general chiral invariant Lagrangian at order $p^4$ in
standard CHPT has been
given by Gasser and Leutwyler \cite{gl85}. The extension to the
graded symmetry version is not needed here, since we are not going
beyond order $p^4$, and are not interested in having the spurious
degrees of freedom as external particles\footnote{Moreover we will not
  consider singlet fields as external particles. They require at least 
  two more counterterms as shown by Eq. (\ref{ZZ}). }: we can use the
standard CHPT Lagrangian right away. 

There is however a slight modification that we have to introduce. As
we noted before, the trace of $\chi_+$ starts with a constant term
proportional to $N$ in the degenerate mass case we are considering here. 
In the quenched version a linear dependence upon
$N$ is forbidden, and therefore we must always substitute $\langle
\chi_+ \rangle \rightarrow \langle \hat\chi_+ \rangle$.
Apart from this modification, we have followed existing notations 
for the choice of the $O(p^4)$ Lagrangian, both in the  $SU(3)$ and $SU(2)$ 
case. The $SU(3)$ choice is the standard Gasser and Leutwyler
Lagrangian \cite{gl85}, while for $SU(2)$ we choose to use the
Gasser--Sainio--\v Svarc Lagrangian \cite{gss}.  

An important point concerns the value of the counterterms: we observe that
in the quenched case the counterterms do not depend on the number of
flavours. Not only the divergent part, as we have explicitly shown in the
previous section, but also the numerical value of the finite part of the
counterterms does not change for different values of $N$.
Therefore it is useful to identify, and give names to them in the
general $N$ case. 
For the more interesting cases of $N=3$ and $N=2$, because of trace
relations, one will be able to access only certain combinations of them, as
we will specify below.
For general $N$ the lagrangian at order $p^4$ is given by:
\be
{\cal L}_4=\sum_{i=0}^{10} \Lambda^q_i P_i \; \; ,
\ee
where the eleven operators $P_i$ are listed in Table \ref{tab:L4} (we
remind the reader that in the quenched case it is necessary to change
$\langle \chi_+ \rangle \rightarrow \langle \hat\chi_+ \rangle$). 
These eleven chiral invariant operators contain, besides those defined in 
Eq. (\ref{DEFF}), the following new building blocks:
\ba
f_{\pm\mu\nu}&=& ul_{\mu\nu}u^\dagger \pm u^\dagger r_{\mu\nu} u
    \; \; , \nonumber\\
\chi_{-}&=& u^\dagger\chi u^\dagger - u\chi^\dagger u  \; \; .
\ea
To derive the results shown in Table \ref{tab:L4} the following
relation is useful:
\be
f_{+\mu\nu}= 2i\, \Gamma_{\mu\nu} -{i\over 2} [u_\mu ,u_\nu ] \; \; ,
\ee
and the identification $\langle\chi_+^2\rangle = 1/2\, \langle
\chi_+^2 +\chi_-^2\rangle$ and $\langle f_+^2\rangle = 1/2\, \langle
f_+^2-f_-^2\rangle$ can be done up to contact terms which contain
external sources only.

\begin{table}
\begin{center}
\begin{tabular}{|c|c|c|c|c|c|}
\hline
 &&\multicolumn{2}{c|}{}&&\\
&$P_i$ &\multicolumn{2}{c|}{Coeff. of $-\frac{1}{(4\pi)^2(d-4)}$}&  &  \\
$i$ &for $SU(N)$&\multicolumn{2}{c|}{}&$SU(3)$&$SU(2)$\\
 &&\multicolumn{2}{c|}{~~CHPT~~~~~~~~~qCHPT~~} && \\
&&\multicolumn{2}{c|}{\phantom{~~CHPT}~~~~~~~~$\langle \chi_+ \rangle
  \to \langle \hat{\chi}_+ \rangle$
}
&&\\
\hline
 &&&&&\\
0&$\langle u_\mu u_\nu u^\mu u^\nu \rangle$ &$\frac{N}{48}$&0&
Eq. (\protect{\ref{CH1}}) & Eq. (\protect{\ref{CH2}})\\ 
 &&&&&\\
1&$\langle u_\mu u^\mu \rangle^2$ &$\frac{1}{16}$&$\frac{1}{16}$&$L_1$ &
$\frac{1}{4}l_1$\\ 
 &&&&&\\
2&$\langle u_\mu u_\nu\rangle\langle u^\mu
u^\nu\rangle$&$\frac{1}{8}$&$\frac{1}{8}$&$L_2$ &$\frac{1}{4}l_2$ \\ 
 &&&&&\\
3&$\langle u_\mu u^\mu u_\nu u^\nu \rangle$ &$\frac{N}{24}$&0&$L_3$&
Eq. (\protect{\ref{CH2}}) \\
 &&&&&\\
4&$\langle u_\mu u^\mu\rangle\langle \chi_+\rangle$
&$\frac{1}{8}$&$\frac{1}{8}-{v_1 \over 2}$&$L_4$& $ {1\over 8} l_4$ \\ 
 &&&&&\\
5&$\langle u_\mu u^\mu \chi_+ \rangle$ &$\frac{N}{8}$&0&$L_5$
&Eq. (\protect{\ref{CH2}}) \\  
 &&&&&\\
6&$\langle \chi_+ \rangle^2$
&$\frac{1}{16}+\frac{1}{8N^2}$&$\frac{1}{16}-{v_2\over 2}+
\frac{\alpha^2}{72} $&  
$L_6$ & $\frac{1}{16} l_3$\\ 
 &&&&&\\
7&$\langle \chi_- \rangle^2$ &0&0&$L_7$ & $-\frac{1}{16}l_7$\\
 &&&&&\\
8&$ \frac{1}{2} \langle \chi_+^2+ \chi_-^2
\rangle$&$\frac{N}{16}-\frac{1}{4N}$  &$-\frac{\alpha}{12}$
&$L_8$ & Eq. (\protect{\ref{CH2}}) \\ 
 &&&&&\\
9&$ -i \langle f_+^{\mu \nu} u_\mu u_\nu \rangle $ &$\frac{N}{12}$&0
&$L_9$ & 
$-\frac{1}{2} l_6$ \\
 &&&&&\\
10&$\frac{1}{4} \langle f_+^2 - f_-^2\rangle $
&$-\frac{N}{12}$&0&$L_{10}$ & $l_5$\\ 
 &&&&&\\
\hline
\end{tabular}
\end{center}
\protect\caption{List of terms of order $p^4$ for $N$ generic, $N=3$
  and $N=2$. In the second and third columns we give the coefficient
  of the divergence coming from the one loop in the standard and
  quenched CHPT case. As we have indicated in the table, the invariants
  containing $\langle \chi_+ \rangle$ have to be changed with $\langle
  \chi_+ \rangle \to \langle \hat{\chi}_+ \rangle$ in the
  quenched case.}   
\label{tab:L4}
\end{table}

\subsection{$SU(3)$ Lagrangian at order $p^4$}
For $N=3$ the Lagrangian at order $p^4$ reads as follows:
\be
{\cal L}_4^{(N=3)}= \sum_{i=1}^{10} L_i^q P_i \; \; ,
\ee
where the operators $P_i$ are defined in Table \ref{tab:L4}.
The $P_0$ operator is linearly dependent from the others through the
following trace relation: 
\be
P_0=\frac{1}{2}P_1+P_2-2P_3 \; \; ,
\label{CH1}
\ee
which implies:
\be
L_1^q=\Lambda_1^q+{1\over 2} \Lambda_0^q \;, \; \; \; 
L_2^q=\Lambda_2^q+ \Lambda_0^q \;, \; \; \; 
L_3^q=\Lambda_3^q- 2 \Lambda_0^q \;, \; \; \; 
\ee
In order to reabsorb the divergences at one loop we define the $L_i^q$
in the following manner:
\ba
L_i^q &=& L_i^{q \; r}(\mu )+ \Gamma_i^q \lambda \; \; , \nonumber \\
\lambda &=& \frac{\mu^{d-4}}{16
  \pi^2}\left[\frac{1}{d-4}-\frac{1}{2}\left( \ln 4\pi
    +\Gamma^\prime(1)+1\right) \right] \; \; ,
\label{lbar3}
\ea
with $\mu$ the renormalization scale, $\lambda$ contains the
divergence at $d=4$ and the coefficients $\Gamma_i^q$ are given by 
\ba
&&\Gamma_1^q=\frac{1}{16}\; \; ,~~~~~~~\Gamma_2^q=\frac{1}{8}\; \; ,
~~~~~~~~ \Gamma_4^q=\frac{1}{8}(1-4v_1)\; \; , \nonumber\\
&&\Gamma_6^q=\frac{1}{16}\left( 1-8v_2+\frac{2}{9}\alpha^2\right) \;
\; , ~~~~~~~\Gamma_8=-\frac{\alpha}{12} \; \; ,
\label{gamma3}
\ea
all the other $\Gamma_i^q$ are zero.

\subsection{$SU(2)$ Lagrangian at order $p^4$}

For $N=2$ the Lagrangian at order $p^4$ reads as follows:
\be
{\cal L}_4^{(N=2)}= \sum_{i=1}^{7} l_i^q Q_i \; \; ,
\label{Lctr}
\ee
where
\ba
Q_1&=& \frac{1}{4} \langle u_\mu u^\mu \rangle^2 \; \; , \nonumber \\
Q_2&=& \frac{1}{4} \langle u_\mu u_\nu \rangle \langle u^\mu u^\nu
\rangle \; \; , \nonumber \\
Q_3&=& \frac{1}{16} \langle \hat\chi_+\rangle^2 \; \; , \nonumber \\ 
Q_4 &=& \frac{1}{8} \langle u_\mu u^\mu \rangle \langle
\hat\chi_+\rangle \; \; , \nonumber \\
Q_5 &=&\frac{1}{4} \langle f_+^2-f_-^2 \rangle \; \; , \nonumber \\
Q_6 &=& \frac{i}{2} \langle f_+^{\mu \nu} u_\mu u_\nu \rangle
\; \; , \nonumber \\
Q_7 &=& -\frac{1}{16} \langle \chi_-\rangle^2 \; \; .
\label{SU2}
\ea
To reduce the number of chiral invariants needed we have used the
following relations:
\be
P_0=-{1\over2} P_1+P_2 \; \; , ~~~~ P_3 = {1 \over 2} P_1 \; \; , ~~~~
P_5 = {1 \over  2} P_4 \; \; , ~~~~ P_8 =
\frac{1}{2} \left( P_6+P_7 \right) \; \;  .
\label{CH2}
\ee
which imply the following relations between the $N=3$ and $N=2$
counterterms: 
\ba
 {1 \over 4} l_1^q=L_1^q+{1 \over 2} L_3^q \; \; , ~~
&{1 \over 4} l_2^q=L_2^q \; \; ,
&{1 \over 16} l_3^q= L_6^q + {1 \over 2} L_8^q \; \; , \nonumber \\
{1 \over 8} l_4^q = L_4^q + {1 \over 2} L_5^q \; \; , ~~~~
&-{1 \over 16} l_7^q = L_7^q+ {1 \over 2} L_8^q \; \; .&
\ea
Note that the trace relations have been written down using the
invariant $\langle \chi_+ \rangle$, and must be reexpressed in terms of
$\langle \hat{\chi}_+ \rangle$ in the quenched case.
This generates a correction to the constants appearing in the ${\cal L}_2$
Lagrangian, see below.
In order to reabsorb the divergences at one loop we define the $l_i^q$
in the following manner:
\be
l_i^q = l_i^{q \; r}(\mu )+ \gamma_i^q \lambda \; \; ,
\ee
with:
\ba
&&\gamma_1^q=\frac{1}{4} \; \; ,~~~~~~~~\gamma_2^q=\frac{1}{2} \; \;
,~~~~~~~~ \gamma_3^q=1-8v_2-\frac{2}{3}\alpha+\frac{2}{9}\alpha^2 \;
\; ,\nonumber\\ 
&&~~~\gamma_4^q= 1-4v_1 \; \; ,~~~~~~~~\gamma_7^q = \frac{2}{3} \alpha
\; \; ,
\label{gamma}
\ea
all the other $\gamma_i^q$ are zero.
We find useful for the analysis of the phenomenology to introduce the
scale independent constants $\overline{l}_i^q$, defined as follows:
\be
\overline{l}_i^q=\frac{32 \pi^2}{\gamma_i^q}l_i^{q \; r}(\mu )
-\ln\frac{M^2}{\mu^2} \; \; .
\label{lbar}
\ee
As we mentioned above, a complete renormalization at the one--loop
level requires, in the quenched case, the renormalization of the
order $p^2$ constant $B_0$ due to divergences proportional to
$m_0^2$. 
In the present case, $\langle\chi_+^2\rangle$ has been eliminated
with the use of the Cayley--Hamilton relations (\ref{CH2}) in favour of
$\langle\hat{\chi}_+\rangle^2$ and $M^2\langle\chi_+\rangle$. The
divergence proportional to the latter can also be reabsorbed in the
renormalization of the $B_0$ parameter. 
Since $P_5=4 Q_4 +2 M^2 \langle u_\mu
u^\mu \rangle$ the constant $F^2$ receives a finite
correction proportional to $L_5^q$.
For later convenience, we define here the
renormalized constants at order $p^2$ in the two--flavour case, in such a
way that they include also finite corrections:
\ba
\bar{\cal L}_2&=& \frac{\bar{F}^2}{4} \langle u_\mu u^\mu + \bar{\chi}_+
  \rangle \; \; , \nonumber \\
\bar{F}^{\tiny{N=2}}&=&F\left( 1+ 4 L_5^q {M^2 \over F^2} \right) \; \; ,
\nonumber \\ 
\bar{B}_0^{\tiny{N=2}}&=& B_0\left[1-{(m_0^2-2\alpha M^2)\over 48\pi^2F^2}
  \left(\ln {M^2\over 
    \mu^2}+1 \right) \right. \nonumber \\
&& \left. ~~~~~~~~ -\left(8 L_5^q+ {\alpha \over 48\pi^2}\right) {M^2
  \over F^2} +b_0(\mu ) \right] 
\; \; ,
\label{B0BAR2}
\ea
where, with an obvious notation $\bar{\chi}_+$ stands for the analogous of
$\chi_+$ which contains $\bar{B_0}$ instead of $B_0$.

\renewcommand{\theequation}{\arabic{section}.\arabic{equation}}
\setcounter{equation}{0}
\section{Analysis of various observables in quenched CHPT}

\label{phen}

In this section we make a complete one--loop analysis of several
observables in quenched CHPT. The main reason for this is to study the
problems generated by quenching in the finite part of the one--loop
corrections, which we have not considered in the generating functional.
As we will see, some of the finite corrections diverge in
the chiral limit. The origin of these divergences can be traced back
to the presence of the double pole in the singlet two--point function. The
double pole carries in the numerator a new mass scale $m_0$ that does
not vanish in the  chiral limit, and hence modifies the chiral power
counting valid in CHPT.  The standard power counting goes as follows:
the chiral order of a generic diagram is given by the simple formula
\be
D_g = 4 L-2 I + \sum_d d N_d \; \; ,
\label{chiral1}
\ee
where $D_g$ is the chiral dimension of a graph $g$ that has $L$ loops,
$I$ internal lines, and $N_d$ vertices of chiral dimension $d$. The
topological relation
\be
L=I-V+1 \; \; ,
\ee
where $V=\sum_d N_d$ is the total number of vertices, can be used to
obtain
\be
D_g = 2 L + \sum_d (d-2) N_d + 2 \; \; .
\label{chiral_order}
\ee
Since in standard CHPT the lowest chiral dimension of a vertex is two, 
the chiral
dimension of a graph is always bigger than two, and increases with the
number of loops and vertices with chiral dimension bigger than two.
In quenched CHPT the situation changes, and we have to allow for the
presence of vertices with chiral order zero, i.e. the
insertions on the singlet propagators proportional to 
$m_0^2$ (that is a constant in the chiral limit).
In this case Eqs. (\ref{chiral1}) through (\ref{chiral_order}) are
still valid,  but due to the presence of terms with $d=0$, $D_g$ may
now be smaller than two, and even negative.
Naively one could conclude that $D_g$ could even be unbounded from
below. However, one has to take into account the fact that virtual
quark loops are forbidden: this puts a series of constraints on the
type of graphs 
with $m_0^2$ insertions that are allowed. For example: it is not
possible for two $m_0^2$ vertices to lie on the same line one after
the other, or, no standard vertices can have all the outgoing
lines that end up on an $m_0^2$ vertex.\footnote{There is one
  exception to this, given by vertices with physical external
  sources. In this case disconnected quark loops are allowed, since
  they are not generated by the QCD determinant (see Section
  \ref{three}).} 
These constraints are such that $D_g$ comes out to be bounded from
below, although it may be negative. The value of the lower bound
depends on the observable -- we will see explicit examples below.

In what follows we are going to analyze: the quark condensate, the
pion mass and decay constant, the vector and 
scalar form factors of the pion and the $\pi\pi$ scattering
amplitude. Although these quantities (with the exception of the form
factors) were already analyzed at
the one--loop level in previous works \cite{Sharpe,qCHPT,bg}, we find it 
useful to present them here again, in view of the renormalization that
we have performed at the level of the generating functional, and also of
our definition of the lagrangian at order $p^4$. 
We make the analysis in the case of two light flavours with
degenerate masses.

\subsection{Quark condensate, pion mass and decay constant}
\label{onetwo}

As anticipated in the previous section the renormalized scalar quark 
condensate plays a crucial role among the
quenched observables in the strong sector, since it contains an explicit
dependence upon the quenched chiral logarithms through
the $\bar{B}_0$ parameter (\ref{B0BAR2}) (everywhere in this section we
shall use the $\bar{B}_0$ parameter as defined in Eq. (\ref{B0BAR2}),
dropping the $N=2$ superscript). We shall see later in the case 
of the scalar form factor that all the $\bar{q} q$ matrix elements
share the same feature.
The renormalized scalar density to one loop in the
two--flavour case is given by 
\be
\langle \bar{q}q\rangle_q = -F_\pi^2 \bar{B}_0 \left[1 + O(M^2)
\right] \; \; , 
\label{QQBAR}
\ee
where we have not written down explicitly the standard chiral
corrections of order $M^2$. The problem with these corrections is that
they contain contributions coming from
counterterms of order $p^4$ that contain only external fields (we
have not written them down in the previous section). These
counterterms cannot be determined on a phenomenological basis: their
presence in the expression of the quark condensate reflects the fact
that away from the chiral limit, this quantity cannot be defined
unambiguously. We refer the reader to Ref. \cite{gl84} for a detailed
discussion of this point.
On the other hand, in the chiral limit, where this ambiguity
disappears, the quark condensate diverges due to the quenched chiral
logarithms inside $\bar{B}_0$.

\begin{figure}[th]
\epsfxsize 9 cm
\epsfysize 1.4 cm
\begin{picture}(50,15) \end{picture}
\epsffile{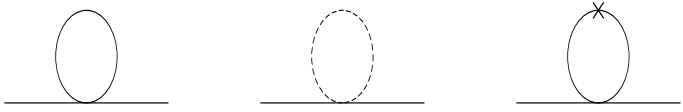}
\protect\caption{One--loop diagrams in quenched CHPT that contribute to
  the squared pion mass $M_\pi^2$. They are the meson tadpole, its ghost
  counterpart and the tadpole with one singlet vertex $(\times )$
  insertion.} 
\protect\label{mpi}
\end{figure}

The pion decay constant to one loop is renormalized only by a finite amount
in the quenched two-flavour case: $F_\pi=\bar{F}$, see
Eq. (\ref{B0BAR2}). Notice that in the quenched three--flavour case there
is no need to define an $\bar{F}$, but on the other hand $L_5^q$ directly
contributes to $F_\pi$ in such a way that for $N=3$ and $N=2$ (as also for
any other $N$) one has the same pion decay constant, as
expected\footnote{We thank Martin L\"uscher for pointing out an
  inconsistency on this point in the first version of the manuscript.}.
The diagrams which renormalize the pion mass to one loop are shown in
Fig. \ref{mpi}. 
The meson tadpole and its ghost counterpart cancel each other, so that the
renormalization of the quenched pion mass at one loop is provided by the 
tadpole with one singlet vertex insertion and its counterterm:
\be
M_\pi^2 = 2\bar{B}_0 m_q \; \;  ,
\label{MASS}
\ee
where $m_q$ is the light quark mass. As one can see, all the one--loop
corrections, including the quenched chiral logarithm have been
reabsorbed in $\bar{B}_0$. Since 
$\bar{B}_0 m_q \sim m_q \log m_q$ when approaching the chiral limit,
the renormalized pion mass tends to zero like $m_q \log m_q $. 
No divergence is produced by quenching in the behaviour of the
renormalized pion mass in the chiral limit, although the way it
approaches zero is different from that of standard CHPT. 
Once $M_\pi^2$ is fixed to its physical value no residual quenched
chiral logarithms are left in the strong sector (with the mentioned
exception 
of $\bar{q}q$ matrix elements). In Appendix \ref{NWI} it is
shown that the same situation occurs in the weak $\Delta I=1/2$
sector, where additional 
quenched chiral logarithms can be reabsorbed in the renormalization of
the weak mass term.

\subsection{Scalar form factor}
\label{three}

\begin{figure}[ht]
\epsfxsize 11 cm
\epsfysize 9 cm
\begin{picture}(50,15) \end{picture}
\epsffile{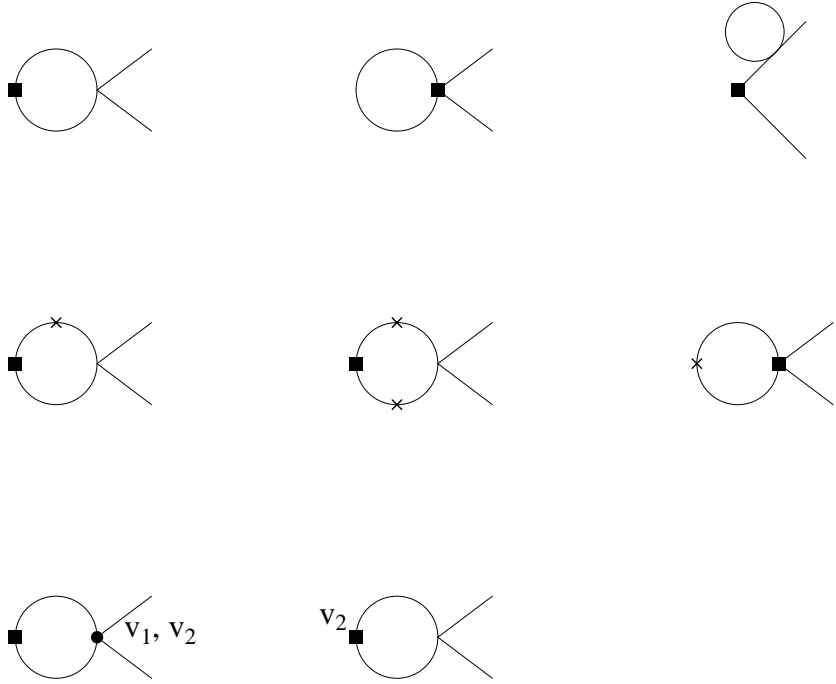}
\protect\caption{One--loop diagrams in quenched CHPT which contribute to
  $F_S(t)$ (the box stands for the scalar source insertion). . They
  are the ``standard'' meson loop diagrams (first line) to 
  which one has to add the corresponding fermionic ghost loop
  diagrams, the singlet insertion diagrams (second line) and the
  diagrams with $v_1$, $v_2$ vertex insertions (third line).}
\protect\label{ffs}
\end{figure}

The scalar form factor of the pion is defined by the matrix element
of the $\bar{q}q$ density between two pion states
\be
\langle \pi^i(p^\prime )|\bar{q}q|\pi^k(p)\rangle = \delta^{ik} F_S(t)
\; \; ,
\label{SM}
\ee
where $t=(p-p')^2$.
In quenched CHPT the complete list of one--loop diagrams
which give contribution to $F^q_S(t)$ are shown in Fig. \ref{ffs}. 

\begin{figure}[ht]
\epsfxsize 13 cm
\epsfysize 2 cm
\begin{picture}(50,15) \end{picture}
\epsffile{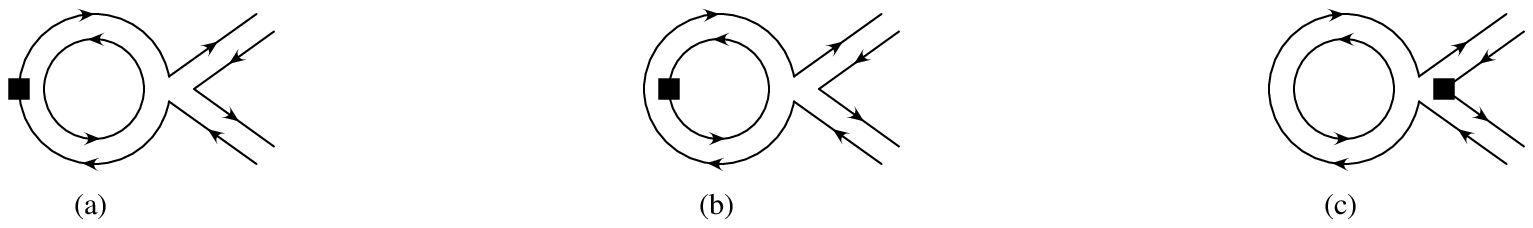}
\protect\caption{One--loop diagrams which contribute to
  $F_S(t)$ in the quark--flow diagram picture. Diagram $(b)$ can be
  present in the quenched approximation, while the others disappear.}
\protect\label{ffsquark}
\end{figure}

An explicit calculation shows that the fermionic ghost loops 
 do not fully cancel the corresponding meson loop diagrams. The reason
 for this mismatch is best understood within the quark--flow diagram
 picture. Here, the physical scalar source only couples
to the quark lines and not to the ghost lines.
The possible one--loop diagrams are the ones listed in Fig. \ref{ffsquark}.
Diagram $(b)$, where the scalar source is coupled to the internal
disconnected closed quark line, has no correspondent ghost loop
diagram: 
this is correct because the loop is not produced by the fermionic
determinant, and must therefore be present also in the quenched
approximation. 

The complete renormalized quenched scalar form factor can be written
as follows
\ba
{F}^q_S(t) &=&F_S^q(0)\left\{ 1+{\bar{J}(t)\over F_\pi^2}\left [ {1\over 2}
\gamma_4^q \left ( t-2M_\pi^2\right )
  + \gamma_3^q M_\pi^2\right ]
+t{\gamma_4^q \over 32\pi^2 F_\pi^2}(\bar{l}_4^q-1)
\right .  \nonumber\\
&&-{2\over 3}{M_\pi^2\over F_\pi^2} \bar{I}_1(t)(m_0^2-\alpha M_\pi^2)
\left ( 1-{2\over 3}\alpha \right ) 
\nonumber\\
&&\left .
+{2\over 9} 
{M_\pi^2\over F_\pi^2} \bar{I}_2(t)(m_0^2-\alpha M_\pi^2)^2
\right\} +O(t^2) \; \;  ,
\ea
where the coefficients $\gamma_i^q$ have been defined in
Eq. (\ref{gamma}) and $F_S^q(0)$ is given by
\ba
{F}^q_S(0) &=&2\bar{B}_0\left\{ 1
-{ (m_0^2-\alpha M_\pi^2)\over 48\pi^2F_\pi^2}
\left ( 1-{2\over 3}\alpha \right ) +{1\over 9} 
{  (m_0^2-\alpha M_\pi^2)^2  \over 48\pi^2 F_\pi^2 M_\pi^2}   
\right .
\nonumber\\
&&\left .  
+{M_\pi^2\over 16\pi^2F_\pi^2}\left [ 
 \gamma_3^q\,(\bar{l}_3^q-1) -  \gamma_4^q\,(\bar{l}_4^q-1) \right
]\right\} \; \; . 
\label{FS0}
\ea
Note that we are working in the degenerate mass case, so that no isospin
breaking effect has been taken into account. In standard CHPT there is no
isospin breaking correction to the scalar form factor at this order of 
the expansion. 
In passing, we state that also in the quenched case
there is no isospin breaking contribution linear in $m_u-m_d$  
to the pion scalar form factor, as it happens in CHPT, while an isospin
breaking correction of order $(m_u-m_d)^2$ is produced via the 
$(\phi_0\, ,\phi_3)$ mixing for neutral pions by the chiral invariant
operator $P_7$ in Table \ref{tab:L4}\footnote{Note that also the
  neutral pion mass $M_{\pi^0}^2$ gets next--to--leading corrections
  of order  $O\left ((m_u-m_d)^2\right )$ from $P_7$ in the quenched
  case.}. 
 
The functions $\bar{J}(t)$, $\bar{I}_1(t)$ and $\bar{I}_2(t)$ are
finite and they are defined in Appendix \ref{APP3}.
The two functions $\bar{I}_1(t)$ and $\bar{I}_2(t)$ are peculiar of
quenched CHPT. They will also appear in the $\pi\pi$ scattering
amplitude, where we shall analyze in some details the various
sicknesses of which they suffer. Here we used their low momentum
expansion to define the scalar form factor at $t=0$.

The scalar form factor is a good example to analyze the modifications
produced by the quenched approximation to an observable at the one
loop level. 
First, the pion loops have been only partially cancelled, therefore
the ordinary chiral logarithms  and the one--loop function $\bar{J}(t)$
do appear in the same way as in standard CHPT, but with different
coefficients (these coefficients may even vanish in particular cases,
like $M_\pi$ and $F_\pi$).
Second, quenched chiral logarithms appear at one loop, but they can be
reabsorbed in the renormalization of the ${B}_0$ parameter, as we have
demonstrated in the previous section.
Besides quenched chiral logs, the remaining finite loop corrections arising
from the anomalous singlet sector and proportional to $m_0^2$ are even
more problematic, since they have negative chiral dimension, as anticipated
in the general discussion above.
It is a simple exercise to calculate the chiral dimension of the one
loop diagram with two $m_0^2$ insertions on the two internal singlet lines
(this is the central graph in the second line of
Fig. \ref{ffs}): with respect to the tree level graph this has
chiral dimension $-2$. These corrections diverge in the chiral limit
like an inverse pion mass squared, see Eq. (\ref{FS0}).
In fact, there is an infinite series of graphs that has the same chiral
dimension: these graphs are obtained from this one by adding any even
number of singlet lines (each one with one $m_0^2$ insertion) between
the two vertices.
Also the insertion of tadpoles and sunset diagrams with the
maximum allowed number of $m_0^2$ insertions does not change the chiral
dimension of the starting diagram.
As far as we could see this series of graphs is also
the one with the lowest chiral dimension for the scalar form
factor. This example shows that despite the general formula
(\ref{chiral_order}) with $d=0$ vertices allowed, in the quenched case
the chiral dimension of amplitudes is bounded from below.

It is also interesting to look at the slope of the scalar form factor
at low momenta in the quenched case. This defines the scalar radius as
follows 
\be
F_S^q(t)=F_S^q(0)\left [ 1+{t\over 6}\langle
  r^2\rangle_S^q+O(t^2)\right ]
\; \; .
\ee
The scalar radius in the quenched approximation at one loop 
is given by: 
\ba
 \langle r^2\rangle^q_S&=& {1\over 16\pi^2 F_\pi^2}\left [  \gamma_3^q
  + \gamma_4^q\, (3 \bar{l}_4^q-4) 
  -\left( 1-{2\over 3}\alpha \right)
  {(m_0^2-\alpha M_\pi^2) \over 3 M_\pi^2}
\right .\nonumber\\
&&\left . ~~~~~~~~~~~~~~~~~~~ 
+ {4\over 45}{(m_0^2-\alpha M_\pi^2)^2\over M_\pi^4}   \right] \; \; .
\ea
In standard CHPT the scalar radius diverges in the chiral limit
because of the presence of $t$ dependent chiral logarithms. It behaves
like: 
\be
 \langle r^2\rangle_S = -{3\over 8\pi^2 F_\pi^2}\,\ln M_\pi^2+\ldots
 \; \;  .
\ee
In the chiral limit the one--loop contribution to the quenched scalar 
radius diverges not just logarithmically as
in the standard case, but like an inverse power of the pion mass:
\be 
 \langle r^2\rangle^q_S\vert_{M_\pi\to 0} \sim {1\over 16\pi^2
   F_\pi^2}\left [ {4\over 45}{m_0^4\over M_\pi^4}-{1\over 3}\left (
     1-{2\over 15}\alpha\right )  
{m_0^2\over M_\pi^2} 
-3\gamma_4^q \log M_\pi^2\right ] + \ldots \; \; .
\ee
The origin of this power--like divergence in the chiral limit is the
same as that of the form factor at $t=0$. Here it is more severe
simply because the definition of the radius implies a derivative with
respect to $t$.

It is interesting to note that in quenched CHPT the Feynman--Hellman
theorem \cite{gl84,FH} does not hold:
\be
{F}^q_S(0) \neq {\partial M_\pi^2\over \partial m_q} \; \; ,
\ee
as one can easily verify by comparing Eq. (\ref{MASS}) and
Eq. (\ref{FS0}). The origin of the violation of this theorem is in
the presence of diagram $(b)$ of Fig. \ref{ffsquark} in the quenched
scalar form factor. This graph cannot be obtained taking a derivative
with respect to $m_q$ of $M_\pi^2$, since the quark loop is not present
in $M_\pi^2$ and cannot be resurrected by a derivative.

\subsection{Vector form factor}

\begin{figure}[t]
\epsfxsize 11 cm
\epsfysize 6 cm
\begin{picture}(50,15) \end{picture}
\epsffile{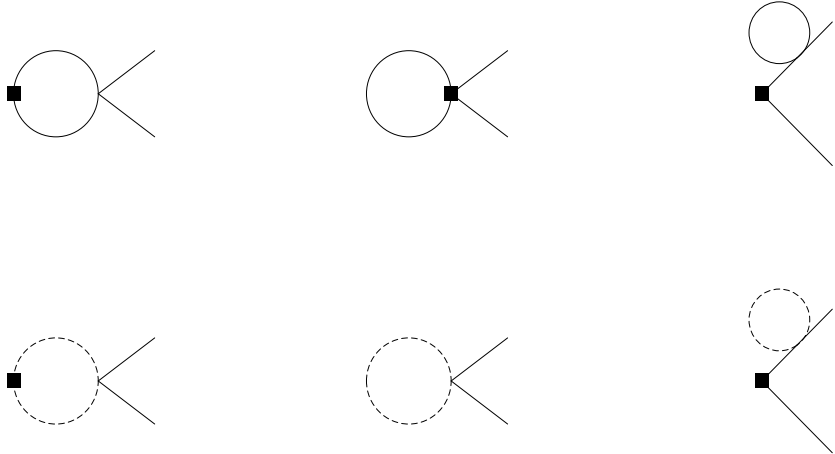}
\protect\caption{One--loop diagrams in quenched CHPT that contribute to
  $F^q_V(t)$ (the box stands for the vector source insertion). They
  are the ``standard'' meson loop diagrams (first line) and the 
  corresponding fermionic ghost loop diagrams (second line). No singlet
  component can run in the loop.}
\protect\label{ffv}
\end{figure}

The vector form factor of the pion is defined in terms of the matrix
element of the vector current $V_\mu^k =
\bar{q}\gamma_\mu{\lambda^k\over 2} q$ between two pion states:
\be
\langle \pi^i(p^\prime )| V_\mu^k|\pi^l(p)\rangle_q =
i\epsilon^{ikl}(p_\mu  
+p_\mu^\prime ) F^q_V(t) \; \; , 
\ee     
where $t=(p-p^\prime)^2$. The divergent contributions to $F^q_V(t)$ 
can be derived from the expression (\ref{ZZ}) of the quenched generating
functional in the usual way.
It is an easy exercise to show that these contributions are zero.
In fact, the only chiral invariant which can give corrections at order 
$p^4$ is the operator number 9 in the list of Table \ref{tab:L4}, 
which has no divergent term in the quenched limit.
In a Feynman diagram approach the graphs which contribute
to one loop are shown in Fig. \ref{ffv}. 
The complete calculation gives zero, because of the systematic
cancellation of each pion loop with the corresponding ghost loop.
In addition, since no singlet component can run in the loop, there is no 
extra contribution coming from the anomalous singlet sector.
The quenched vector form factor for $N=2$ can be written as follows 
\be
F^q_V(t) = 1-{l_6^q\over F_\pi^2} t + O(p^4)\; \; ,
\ee
where the finite counterterm $l_6^q$ is defined in Table \ref{tab:L4}
and Eq. (\ref{SU2}).
Again, no isospin breaking effects have been taken into account. In
standard CHPT the Ademollo--Gatto theorem \cite{AD} guarantees that
they are absent at this order. 
In quenched CHPT the theorem is also valid.  
Note that the counterterm $P_7$ cannot contribute at all to the vector
current matrix element, while the new chiral invariant term $\langle
u_\mu\rangle\langle u^\mu\chi_+\rangle$ induced by the dynamical
singlet component gives $O\left( (m_u-m_d)^2\right )$ corrections to
the decay amplitude $\pi^+\to \pi^0 e\nu$ via the $(\phi_0\, ,\phi_3
)$ mixing. 

\begin{figure}[t]
\epsfxsize 14 cm
\epsfysize 3.2 cm
\epsffile{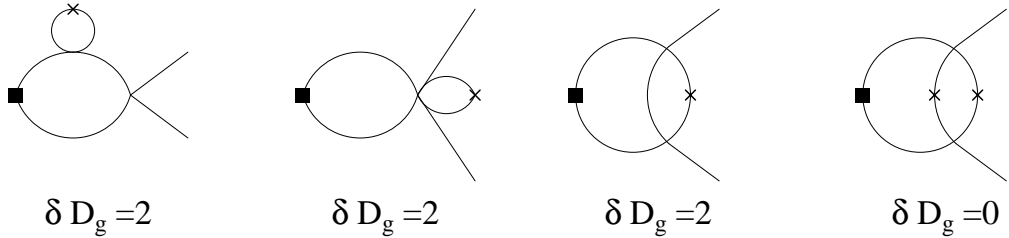}
\protect\caption{Two--loop diagrams in quenched CHPT with the $m_0^2$ 
singlet vertex ($\times$) insertions that give divergent contributions 
to the e.m. charge radius in the chiral limit. They are tadpoles,
which generate quenched chiral logs and the fish diagrams that also
generate power--like divergences. For each diagram the chiral dimension
respect to the tree level is given.} 
\protect\label{ffv2l}
\end{figure}

Since the vector form factor does not receive contributions from
singlets running inside the loop at the one--loop level, it does not show
any divergence in the chiral limit. The situation however changes at two
loops, where we have among others the graphs shown in Fig. \ref{ffv2l}. 
The most dangerous graph is the fish diagram with two $m_0^2$
insertions (the last of Fig. \ref{ffv2l}) which 
has chiral dimension zero respect to the tree level. 
Again this is only the first example of a
full series of graphs which have the same chiral dimension: they are
obtained from the starting one by 
inserting any even number of singlet lines between the same
two vertices as those of the two--loop fish diagram, or tadpoles and
sunset diagrams all with the maximum allowed number of $m_0^2$
insertions. In this case there are no graphs which are more singular
than those in the chiral limit.

The low energy representation of $F^q_V(t)$ also determines the
electromagnetic charge radius of the pion in the quenched approximation
\be
F^q_V(t) = 1+{t\over 6} \langle r^2\rangle^q_V +O(t^2) \; \; .
\ee
In standard CHPT the presence of $t$ dependent chiral logarithms makes the
electromagnetic charge radius diverge in the chiral limit \cite{gl84}
\be
\langle r^2\rangle_V = -{1\over 16\pi^2F_\pi^2} \log M_\pi^2 +\ldots
\; \;  .
\ee
The divergence of the electromagnetic charge radius in full QCD can be
understood in a physically intuitive way.
The charge distribution is cut off by the Yukawa potential $\sim
e^{-M_\pi r}$ at large distances. In the chiral limit $M_\pi$ goes to
zero and  the Yukawa potential is no more effective, the charge
distribution falls off like a power of the distance and the charge radius
becomes infinite. The charge distribution of the
pion cloud surrounding any particle gets modified by quenching. 
As a consequence, the behaviour of the charge radius in the chiral
limit is modified.
In the quenched case the one--loop contribution gives
\be
\langle r^2\rangle^q_V = -{6 l_6^q\over F_\pi^2} \; \; ,
\ee
which stays finite in the chiral limit. 
The situation changes at two loops and higher: the graphs that we have
discussed above, which have chiral dimension zero with respect to the
tree level (like the two--loop fish diagram), do generate power--like
divergences in the chiral limit. 
At two loops we are going to have a behaviour like:
\be
\langle r^2\rangle_V^{q~{\mbox{\tiny{2 loop}}}}\vert_{M_\pi\to 0} \sim
{1\over (16\pi^2 F_\pi^2)^2} \left ( d_1\,{(m_0^2/N_c)^2\over M_\pi^2}
+d_2\, {m_0^2\over N_c}\, \ln M_\pi^2\right ) \; \; ,
\ee  
where presumably also at this order the chiral logs could be reabsorbed
in the renormalization of some order $p^4$ constants.

\subsection{The $\pi\pi$ scattering amplitude}
\label{pion}

The $\pi\pi$ scattering amplitude is another example of an observable
where one can find all the typical effects of quenching.
Moreover it is an interesting quantity by itself since a comparison of
the prediction for the two $S$--wave scattering lengths with existing
lattice calculations \cite{fuku} is possible.
  
\begin{figure}[t]
\epsfxsize 11 cm
\epsfysize 3 cm
\begin{picture}(50,15) \end{picture}
\epsffile{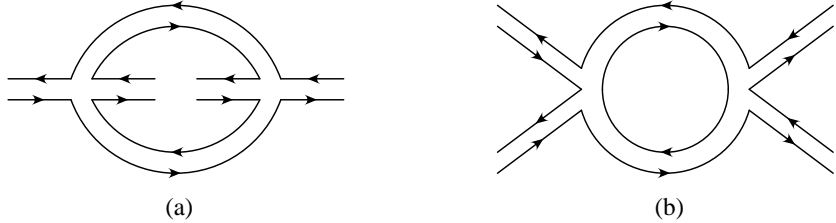}
\protect\caption{Two examples of pion loop graphs contributing to $\pi
  \pi$ scattering in the quark--flow diagram picture (all lines are
  quark lines). Diagram (a) does not contain quark loops, whereas
  diagram (b) does.} 
\label{fig1}
\end{figure}

\begin{figure}[t]
\epsfxsize 14 cm
\epsfysize 11 cm
\begin{picture}(30,15) \end{picture}
\epsffile{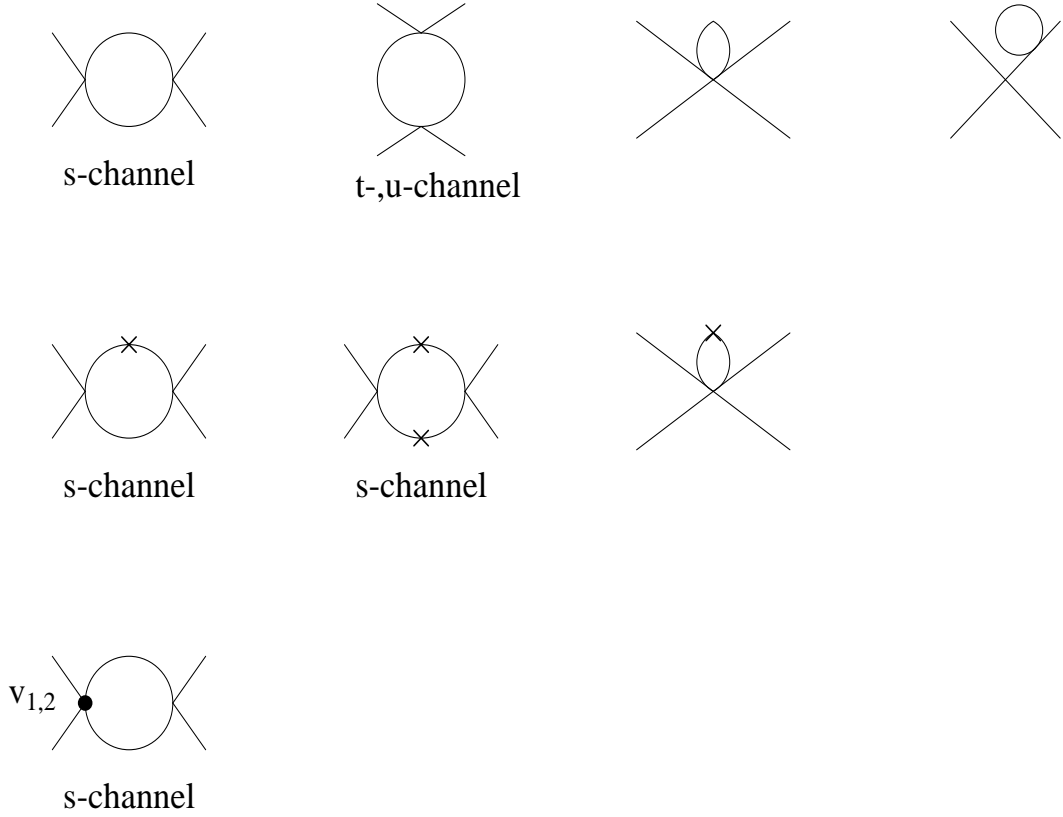}
\protect\caption{One--loop diagrams in quenched CHPT which contribute to
  the $\pi\pi$ scattering amplitude in the two degenerate flavour case. 
They are the ``standard'' meson loop diagrams (first line) to
which one has to add  the corresponding fermionic ghost loop diagrams, 
the singlet vertex $(\times )$ insertion
diagrams (second line) and the diagrams with one $v_1$, $v_2$ 
vertex insertion (third line).}
\protect\label{pipi}
\end{figure}

The presence of ``standard'' chiral logs even in the quenched theory
has to be interpreted as due to diagrams with pion loops that do not
contain quark loops. For the $\pi \pi$ scattering amplitude an example
is given in Fig. \ref{fig1}. 
The one--loop contributions to the $\pi\pi$ scattering amplitude in
quenched CHPT come from the diagrams shown in Fig. \ref{pipi}.   
The scattering amplitude at tree level is the same as in standard CHPT
\be
A^{\mbox{\tiny{tree}}}(s,t,u) = {s-M^2\over F^2} \; \; ,
\ee
where $M$ and $F$ are the bare pion mass and decay constant.
The renormalized scattering amplitude in quenched CHPT and in the 
two degenerate flavour case can be written as follows
\be
A(s,t,u)={s-M_\pi^2\over F_\pi^2}+ B(s,t,u)+C(s,t,u)+O(p^6) \; \; ,
\ee
where
\ba
B(s,t,u)&=&
{\bar{J}(s)\over 4F_\pi^4} \left\{  s^2-16M_\pi^2 
v_1 
(s-2M_\pi^2) +4M_\pi^4\left ( \gamma_3^q -1\right ) \right\}    \nonumber\\
&& +{1\over 4F_\pi^4} \left\{ \bar{J}(t)\left(t-2M_\pi^2\right)^2
+\bar{J}(u) \left(u-2M_\pi^2\right)^2 \right\}   \nonumber\\
&& +I_1(s){2M_\pi^4\over 3F_\pi^4} \left( m_0^2-\alpha M_\pi^2\right)
\left({2\over 3}\alpha -1\right)
+I_2(s){2M_\pi^4\over 9F_\pi^4} \left( m_0^2-\alpha M_\pi^2\right)^2
\; \; ,
\nonumber\\
&&\nonumber\\
C(s,t,u)&=&{1\over 128\pi^2F_\pi^4}\left\{ {\vrule height1.08em
    width0em depth1.08em} 
s^2 (2\bar{l}_1^q+\bar{l}_2^q-3) +(t-u)^2 (\bar{l}_2^q-1) \right
.\nonumber\\ 
&& +8sM_\pi^2\left [1-\bar{l}_1^q + \gamma_4^q\, (\bar{l}_4^q -1)\right ]
\nonumber\\
&&\left . +8 M_\pi^4\left [ \bar{l}_1^q-1+ \gamma_3^q\, (\bar{l}_3^q-1)
 -2 \gamma_4^q\, (\bar{l}_4^q-1)  \right ] 
{\vrule height1.08em width0em depth1.08em} \right\} \; \; .
\ea
For a definition of the functions
$\bar{J}(q^2)$, $I_1(q^2)$ and $I_2(q^2)$ see Appendix
\ref{APP3}. 
The functions $I_1(s)$ and $I_2(s)$ arise from diagrams with one
and two $m_0^2$ insertions on the two internal singlet lines in the
$s$--channel 
respectively (see Fig. \ref{pipi}).
Note that everything is expressed in terms of the renormalized squared
pion mass $M_\pi^2$ given by Eq. (\ref{MASS}) and $F_\pi = \bar{F}$. 
Note also that any dependence upon quenched chiral logarithms has been
again reabsorbed in the $\bar{B}_0$ parameter contained in the
renormalized pion mass, as expected.
The function $C(s,t,u)$ contains only polynomial contributions, while 
the invariant function $B(s,t,u)$ is the quenched analogue of the
unitarity correction to the scattering amplitude in ordinary CHPT. It
is important to note that unitarity is destroyed by the quenched
approximation: the structure of the cuts in the one--loop amplitude is
not related via unitarity to the real part of the tree level
amplitude. 
Moreover, one can easily verify that the Fermi--Watson theorem, which
relates, e.g., the imaginary part of the vector and scalar form
factors to those of the corresponding partial waves of $\pi \pi$
scattering, is not valid in this case.

In this particular example the violation of unitarity is also
immediately seen in the presence of the finite functions  $I_1(q^2)$
and $I_2(q^2)$, which are not generated in ordinary CHPT. They have 
a nonzero imaginary part for $s\geq 4M^2$ that has a singularity at
$s=4 M^2$ (of the type $(s-4M^2)^{-1/2}$ and $(s-4M^2)^{-3/2}$,
respectively, see Appendix \ref{APP3}), which is a pure quenching
artifact. These singularities have been already identified in
\cite{bg,pipiq}. Here we have rederived them in the $\alpha\neq 0$
case and inserted in the complete formula for the amplitude.

Interesting quantities to be extracted from the $\pi\pi$ scattering
amplitude are the $S$--wave scattering lengths. In Ref. \cite{pl} we
calculated the coefficients of the chiral logarithms which arise in
the quenched case and made the comparison with standard CHPT. Here we
give the complete expression of the $S$--wave scattering lengths in
the isospin $I=0,2$ channels to one loop and comment on the anomalous
behaviour of the isospin amplitude in the $I=0$ channel (which was
already remarked in Ref. \cite{bg}). The $I=0,2$ amplitudes are
expressed in terms of the invariant amplitude $A(s,t,u)$ as follows
\ba
T^0(s,t)&=&3A(s,t,u)+A(t,u,s)+A(u,s,t) \; \; , \nonumber\\
T^2(s,t)&=&A(t,u,s)+A(u,s,t) \; \; . 
\ea
The pion scattering lengths $a^I_l$ for a given isospin $I$ and
angular momentum $l$ are defined by the behaviour of the
partial wave amplitudes near threshold
\be
Re ~t^I_l(s) = q^{2l} \left\{ a^I_l +q^2 b_I^l +O(q^4)\right\} \; \; ,
\ee
which enter the expansion in partial waves of the isospin amplitude
\be
T^I(s,t)= 32\pi \sum_{l=0}^\infty (2l+1) P_l(\cos\theta ) t^I_l(s) \;
\; .
\ee
For more details about the notation we refer the reader to
Ref. \cite{gl84}. 
The scattering amplitude in the $I=0$ channel contains the amplitude
$A(s,t,u)$ and therefore 
in the quenched case acquires a sick threshold behaviour due to the
presence of functions $I_1(s)$ and $I_2(s)$. These functions do not
contribute to the $I=2$ amplitude.
On the other hand the divergences at threshold present in the infinite
volume case show up as ``enhanced'' finite volume corrections to the
L\"uscher formula \cite{luescher}, that is used on the lattice to
extract the scattering lengths; these finite volume corrections have
been studied in Ref. \cite{bg}. 
We can formally define the quenched $I=0$ 
$S$--wave scattering length $a^0_0$ as the
coefficient of the $(q^2)^0$ term in the expansion of the real part of 
the isospin amplitude $T^0(s,t)$ in partial waves. This gives us an
idea of the size of normal one--loop corrections to the scattering length. 
The present definition is also equivalent to the one adopted in
Ref. \cite{bg} in the analysis of the finite volume corrections. 
The quenched $S$--wave scattering length in the $I=2$ channel $a^2_0$
is defined in the usual way. 
For the complete renormalized $S$--wave ``quenched scattering
lengths'' at one loop we find:
\ba
{32\pi F_\pi^2 \over M_\pi^2}  a_0^0&=& 7
 + {M_\pi^2\over 16\pi^2 F_\pi^2}\left\{ 7+ 5(\bar{l}_1^q+2\bar{l}_2^q) 
+\gamma_3^q \,(5\bar{l}_3^q+1)+2\gamma_4^q \,(\bar{l}_4^q-1)  -48 v_1
\right\} 
\nonumber\\
&&-{(m_0^2-\alpha M_\pi^2)\over 
48\pi^2F_\pi^2}\left ({2\over 3}\alpha -1\right )
+{5\over 9}{(m_0^2-\alpha M_\pi^2)^2\over 48\pi^2 M_\pi^2F_\pi^2} \;
\; ,
\label{a00}
\ea
\ba
{32\pi F_\pi^2 \over M_\pi^2} a_0^2&=&  -2
+ {M_\pi^2\over 16\pi^2 F_\pi^2}\left\{  
 2(\bar{l}_1^q+2\bar{l}_2^q -1) +2\gamma_3^q \, (\bar{l}_3^q-1)
 -4\gamma_4^q \, (\bar{l}_4^q-1) \right\}\nonumber\\
&&+{(m_0^2-\alpha M_\pi^2)\over 
24\pi^2F_\pi^2}\left ({2\over 3}\alpha -1\right )
+{2\over 9}{(m_0^2-\alpha M_\pi^2)^2\over 48\pi^2 M_\pi^2F_\pi^2}
\; .
\label{a02}
\ea
The renormalized quenched scattering lengths depend upon four
counterterms $\bar{l}_1^q,\ldots\bar{l}_4^q$ and the parameters of the
anomalous singlet sector at leading order, $m_0^2$, $\alpha$, $v_1$ and
$v_2$. The counterterms $\bar{l}_i^q$ carry the chiral logarithms
$\bar{l}_i^q=-\log m +\ldots $. In ordinary CHPT the chiral logarithms
are largely dominant in the one--loop corrections to the $S$--wave
scattering lengths at the renormalization scale $\mu =1$ GeV
\cite{gl83}.  Here the main unknown is the value of the parameters
$v_1,v_2$ of the singlet sector.
The singlet parameters $m_0$ and $\alpha$ can be extracted from lattice
calculations. Favoured values are listed e.g. in Ref. \cite{lat96}. 
With these values at hand we can do the following numerical
exercise. Let us disregard for the moment the parameters $v_1$ and
$v_2$ and limit the analysis to the contributions that are reasonably
expected to be the dominant ones: 1) the singlet corrections in $m_0$
and $\alpha$ and 2) the standard chiral logarithms.
With the definitions 
\be
\delta = {m_0^2\over 48\pi^2 F_\pi^2} \; \; ,~~~~~\epsilon = 
{M_\pi^2\over 48\pi^2 F_\pi^2} \; \; ,~~~~\bar{\delta} =
{m_0^2-\alpha M_\pi^2 \over 48\pi^2 F_\pi^2}=\delta-\alpha\epsilon \;
\; ,
\ee
the leading contributions to the scattering lengths are as follows:
\ba
{32\pi F_\pi^2 \over M_\pi^2}  a_0^{0}&=& 7 -\left( {2\over 3}\alpha
  -1\right) 
 \bar{\delta} +{5\over 9}{\bar{\delta}^2 \over \epsilon}
- 66 \epsilon \ln{M_\pi^2\over \mu^2} + \ldots \;\; , \nonumber \\ 
{32\pi F_\pi^2 \over M_\pi^2}  a_0^{2}&=& -2 + \left( {2\over 3}\alpha -1
\right) 2\bar{\delta} +{2\over 9}  {\bar{\delta}^2 \over \epsilon}
- 12 \epsilon \ln {M_\pi^2\over \mu^2} + \ldots \;\; .
\label{a0leading}
\ea
For the numerical calculations we use $F_\pi = 93$ MeV,
$\delta = 0.15$ and $\alpha = 0.6$ and vary the pion mass between its
physical value $M_\pi= 140$ MeV, and $M_\pi=600$ MeV, which is
presumably already outside a reasonable range of validity for ordinary
CHPT. The chiral log is evaluated at $\mu=1$ GeV.
The numerical results are given in Tables \ref{taba00} and
\ref{taba02} for the $I=0$ and $I=2$ scattering lengths
respectively. We note that at the physical value of the pion mass the
$\bar{\delta}^2 / \epsilon$ term is largely dominant in both cases:
the divergence in the chiral limit produced by quenching is already
felt at the physical pion mass. This also means that the whole
framework is not very reliable in this range, since also higher loop
effects may produce modifications of the same chiral order (higher
powers of $\delta$ with the same $1/\epsilon$ in front). 
At larger values of the pion mass, which are those typically used in
lattice calculations, the situation changes and the standard chiral
logarithms become dominant, as it happens in standard CHPT.
\begin{table}[thb]
\begin{center}
\begin{tabular}{|c|c|c|c|c|}
\hline
&&&&\\
$M_\pi$ (MeV) & tree & $\bar{\delta}$ &  $\bar{\delta}^2/\epsilon$ &
$\epsilon \ln M_\pi^2$ \\
\hline
&&&&\\
140 & 7 & 0.09 & 2.5  & 1.2 \\
300 & 7 & 0.08 & 0.47 & 3.5 \\
600 & 7 & 0.06 & 0.06 & 5.9 \\
\hline
\end{tabular}
\end{center}
\protect\caption{Numerical values of the leading contributions to $a_0^0$
  quenched up to one loop for $M_\pi=140, \; 300, \; 600$ MeV,
  according to Eq. (\protect{\ref{a0leading}}).}
\label{taba00}
\end{table}

\begin{table}[thb]
\begin{center}
\begin{tabular}{|c|c|c|c|c|}
\hline
 &&&&\\
$M_\pi$ (MeV) & tree & $\bar{\delta}$ &  $\bar{\delta}^2/\epsilon$ &
$\epsilon \ln M_\pi^2$ \\
\hline 
&&&&\\
140 & -2 & -0.18 & 1.0  & 0.23 \\
300 & -2 & -0.16 & 0.19 & 0.63 \\
600 & -2 & -0.12 & 0.02 & 1.1 \\
\hline
\end{tabular}
\end{center}
\protect\caption{Numerical values of the leading contributions to $a_0^2$
  quenched up to one loop for $M_\pi=140, \; 300, \; 600$ MeV
  according to Eq. (\protect\ref{a0leading})}   
\label{taba02}
\end{table}

This picture, although at a semiquantitative level,
suggests that quenched lattice calculations of the $S$--wave
scattering lengths with a moderately high pion mass (like the ones in 
Ref. \cite{fuku}), should not be too far from those predicted by full
CHPT. This conclusion is based on two observations:
first the standard chiral logarithms start soon to be dominant with
respect to the dangerous quenching effects, and second
their coefficient happens not to be substantially changed by quenching
\cite{pl}.
The comparison between the standard CHPT prediction at one
\cite{gl84} and two loops \cite{2lpipi}, and the lattice calculation
\cite{fuku}, has been made in Ref. \cite{g}.

\renewcommand{\theequation}{\arabic{section}.\arabic{equation}}
\setcounter{equation}{0}
\section{Summary and conclusions}

In this paper we have analyzed the quenched version of Chiral
Perturbation Theory at the one--loop level.
We have calculated the one--loop ultraviolet divergences of 
the theory at the level
of the generating functional, and shown how one can reabsorb
all those divergences by a proper definition of the counterterms. 
We have shown that even in the presence of the anomalous singlet
sector the ultraviolet 
divergent part of the quenched generating functional can be
calculated in closed form.  
We have closely followed the notation and methods of standard CHPT
\cite{gl84} in order to identify as clearly as possible the changes
produced by the quenched approximation in the formulation of 
the effective theory.

\paragraph{}
We have found a systematic cancellation of the flavour--number
dependent terms inside the divergent part of the generating functional
to one loop. 
As we anticipated in Ref. \cite{pl} the complete $N$--independence of
quenched CHPT is welcome, since it shows that we understand the
differences between standard CHPT and its quenched version.
Let us recall that the calculation of the divergences to one loop in
CHPT produces explicit $N$ dependence, in three different powers: $N$,
$1/N$ and $1/N^2$. The terms linear in $N$ must be generated at the
quark level by virtual quark loops: therefore they must be absent in
the quenched theory. 
The terms with inverse powers of $N$ are generated by the decoupling
of the singlet field from the octet of the Goldstone bosons.
Since the decoupling does not take place in the quenched theory, also 
the inverse powers of $N$ disappear in qCHPT to one loop.
A posteriori one could say that the changes that lead from standard
CHPT to its quenched version could have been guessed by simply looking
at the $N$ dependence of the generating functional to one loop. 
In fact this can still be done in other sectors of the effective
theory that have not been fully analyzed yet. We give one example of
this in Appendix \ref{NWI}, where we study the one--loop divergences in
the sector of the on--shell non--leptonic weak interactions.

\paragraph{}
The quenched approximation produces a double pole in the singlet two--point
function, which however is not allowed in a
consistent quantum field theory, and is therefore the source of many
sicknesses of  quenched CHPT. As was shown already in
Refs. \cite{Sharpe,qCHPT}, one of the consequences of this double
pole is the appearance of a new kind of chiral logarithms in the
one--loop corrections. Together with the standard $M^2 \ln M^2$ chiral
logarithms, qCHPT has corrections of the form $m_0^2 \ln M^2$, which
diverge in the chiral limit. 
The complete calculation of all the  ultraviolet one--loop divergences in the
generating functional, has shown that the quenched chiral logs  
can be accounted for via a renormalization of the lowest order constant
$B_0$  (which is proportional to the quark condensate). As a consequence, 
the renormalized $\bar{B}_0$ parameter diverges in the chiral limit, while the
renormalized pion mass  $M_\pi^2 = 2 \bar{B}_0 m_q$ does not.
 The use of the renormalized pion
mass to express any other observable makes the quenched chiral logs
disappear at one loop, with the only exception of $\bar{q} q$ matrix
elements, that are proportional to the renormalized $\bar{B}_0$
parameter alone. Hence, $\bar{q}q$ matrix elements remain the unique place  
for discovering the presence of quenched chiral logs in quenched
lattice calculations within the strong sector. 

\paragraph{}
The double pole in the singlet two--point function also changes the standard
chiral power counting for which diagrams with a higher number of loops
are of higher chiral order. In the quenched case one may have
graphs with any number of loops with the same chiral dimension, and
the chiral order of an amplitude is no more constrained to be
positive: as a consequence, quenched CHPT has power--like
divergences in the chiral limit. These divergences are in principle a
very serious problem of the theory, although they seem to be
a unavoidable consequence of the quenched approximation. Since the
graphs that have negative chiral dimension are also ultraviolet finite, their
study requires the calculation of the UV finite part of the loop
corrections. We have shown how they arise within the generating functional
approach; at one--loop and at order $\Phi^4$ they are given in
Eq. (\ref{IRdiv}).
We have therefore analyzed some physical quantities at one
loop in the case of two degenerate light flavours: the scalar quark
condensate, the pion mass, the scalar and vector form factors of the
pion and the $\pi\pi$ scattering amplitude. 
This has given us the possibility to discuss in detail the changes 
induced by quenching in the UV finite part of the one--loop
corrections. The main changes can be summarized by saying that
unitarity is not satisfied anymore, and that the double pole in the
singlet two--point function produces singularities in the chiral limit, and
also unphysical singularities at threshold. 

\paragraph{}
The differences between CHPT and its quenched version
are rather well understood, as the study of the flavour--number
dependence of the generating functional at one--loop also shows. The
presence of the double pole in the singlet two--point function is also a
rather direct consequence of the quenched approximation. This double pole
has dramatic effects on the effective theory. However, it looks
plausible that despite all these inconsistencies (or
maybe because of them) quenched CHPT is the right tool to
understand the effects of quenching in actual lattice
calculations. The crucial check will be a detailed comparison of qCHPT
predictions with the quark mass dependence of various quenched
quantities on the lattice, and especially of the way they 
approach the chiral limit.
We expect that further investigations in this direction will answer
these questions.

\section*{Acknowledgements}
We thank Roberto Petronzio and Juerg Gasser for many enlightening
discussions, Gerhard Ecker and Joachim Kambor for informative
discussions about the weak non--leptonic sector. 
This work has been supported by Schweizerisches Nationalfonds and by
the HCM, EEC--Contract No. CHRX--CT920026 (EURODA$\Phi$NE). 

\newpage
\appendix
\renewcommand{\theequation}{\Alph{section}.\arabic{equation}}
\setcounter{equation}{0}
\section{Divergences proportional to $m_0^2$}

\label{m0}

In this appendix we  try to explain in a simple way why the
divergence proportional to $m_0^2$ is given by the chiral invariant
term $\langle \chi_+ \rangle$.
We start from the observation that the double
pole in the singlet propagator can be expressed with a derivative of a
normal propagator:
\be
\frac{m_0^2}{\left(M^2-p^2\right)^2} = -m_0^2 \frac{\partial}{\partial
  M^2} \frac{1}{M^2-p^2} \; \; .
\ee
In fact in the quenched case one is keeping only the first two terms
in a Taylor series:
\be
\frac{1}{M_{\eta'}^2-p^2} =
\frac{1}{M^2-p^2}-(M_{\eta'}^2-M^2)\frac{1}{\left(M^2-p^2 \right)^2} +
\ldots \; \; ,
\ee
where $m_0^2=M_{\eta'}^2-M^2$. The divergences proportional to $m_0^2$
arise from this Taylor expansion of the propagator.
Before the expansion, singlet loops give two type of divergent
contributions:
\be
{1 \over i} \int { d^d p \over (2 \pi)^d} {1 \over
  M_{\eta^\prime}^2-p^2} 
\; \; , \; \; \mbox{and} \; \; \; \; \;  {1 \over i} \int {d^d p \over
  (2 \pi)^d} {1 
\over M_{\eta^\prime}^2-p^2} {1 \over M_{\eta^\prime}^2-(p-q)^2} \; \; . 
\ee
The divergence proportional to $m_0^2$ is obtained expanding the
singlet propagator inside the loop integrals and taking only the
second term in the expansion.
It is clear that only the tadpole produces a divergence:
\be
{1 \over i} \int {d^d p \over (2 \pi)^d} {1 \over M_{\eta^\prime}^2-p^2}
\rightarrow {1 \over i} \int {d^d p \over (2 \pi)^d} {-m_0^2 \over \left(
  M^2-p^2\right)^2} = -m_0^2 J(0) \; \;.
\ee
The chiral structure of the term proportional to the tadpole is very
easily identified, and is given by the term proportional to $\xi_0^2$
after expanding the action in fluctuations around the classical
solution:
\be
{\cal L}_2 = \frac{F^2}{4} \langle u_\mu u^\mu+\chi_+ \rangle
\rightarrow
\frac{F^2}{4} \left\{ -\xi_0\left( \Box+M_{\eta'}^2 \right)\xi_0 -
  \xi_0^2 \left( \langle \chi_+ \rangle 
    -2NM^2 \right) \right\} \; .
\label{xi0exp}
\ee
Obviously, the tadpole is generated by the contraction of the two
$\xi_0$'s in the last term of Eq. (\ref{xi0exp}), and therefore the
divergence is proportional to $\langle \chi_+ \rangle$.

In summary: the divergence proportional to $m_0^2$ must be, for
dimensional reasons, a chiral term of order $p^2$. This divergence
comes out from a tadpole through a derivative with respect to
$M_{\eta'}^2$. In the tadpole the vertex is a chiral invariant of
order $p^2$. 
The simplified recipe to derive the form of the divergent
term proportional to $m_0^2$ amounts to determining the chiral
invariant that is proportional to $\xi_0^2$ after expanding the CHPT
Lagrangian in fluctuations around the classical solution.

The recipe applies in the same way to other sectors. We will give an
example below for the non--leptonic weak interactions sector.

\renewcommand{\theequation}{\Alph{section}.\arabic{equation}}
\setcounter{equation}{0}
\section{Non--leptonic weak interactions}

\label{NWI}

In this appendix we show how one can derive the structure of the
divergences in the quenched case just by looking at the $N$ dependence
of the divergences of the standard CHPT case.
We will consider the weak octet Lagrangian which contributes to 
the non--leptonic weak interactions with
$\Delta I = 1/2$. The structure of the divergences in standard CHPT
has been given by Kambor, Missimer and Wyler \cite{kmw}, and then
expressed in terms of a minimal basis by Ecker, Kambor and Wyler
\cite{ekw} for on--shell processes. 
We will use the basis given in the latter Reference.

Let us recall here some basic notation.
In this sector the lowest order Lagrangian is given by:
\be
{\cal L}_{W2}^8= c_2 \langle \Delta u_\mu u^\mu \rangle + c_5 \langle
\Delta \chi_+ \rangle \; \; ,
\ee
where $\Delta = u \lambda_6 u^\dagger$, and $c_{2,5}$ are low energy
constants. The $c_5$ term can be omitted for on--shell processes, as it
can be transformed away by a field redefinition. We will not consider
it anymore in what follows.
Since the above--mentioned calculations of the divergences were made for
$N=3$ we have redone the calculation of the divergences for $N$ generic. 
Our result for the divergent part of the one--loop generating
functional, reads as follows:
\ba
Z^8_{\mbox{\tiny{one loop}}}&=& -\frac{\mu^{d-4}}{16 \pi^2}
\frac{1}{d-4} {c_2\over F_\pi^2}\, \int\, dx\, 
L^8_{\mbox{\tiny{div}}} + \mbox{finite terms} \; \; , \nonumber \\
L^8_{\mbox{\tiny{div}}} &=& W_4+\frac{1}{2} W_6-\frac{3}{4} W_7
+\frac{1}{4}W_8 
-\frac{1}{2} W_{11} \nonumber \\
&+& \frac{1}{N} \left[ 2 W_{10}+W_{12}-2 W_{21} -2 W_{22} +W_{36}
  -\frac{2}{N} W_{11} \right] \nonumber \\
&+& N \left[ \frac{2}{3} W_1 -\frac{1}{6} W_2 +\frac{1}{4} W_5
  +\frac{1}{4} W_9 -\frac{1}{4} W_{12} +\frac{1}{12} W_{14}
  +\frac{1}{6} W_{15} - \frac{1}{12} W_{16} \right. \nonumber \\
& &   \; \; \; \; -\frac{1}{24} W_{18}
  -\frac{5}{12} W_{19} +\frac{1}{4} W_{20} +\frac{1}{2} W_{21}
  +\frac{1}{2} W_{22} +\frac{1}{6} W_{25} -\frac{1}{4} W_{26}
  \nonumber \\ 
& & \; \; \; \;\left. + \frac{1}{24} W_{27} -\frac{1}{4} W_{36}
  -\frac{1}{24} 
  W_{37}  -\frac{1}{4} W_{38} \right] \; \; .
\ea
With $W_i$ we have indicated the operators of order $p^4$ given in
Ref. \cite{ekw}. Here we have one more:
\be
W_{38} = \langle \Delta u_\mu \chi_+ u^\mu \rangle \; ,
\ee
that for $N=3$ is linearly dependent on the other operators thanks to
the trace identity coming from the Cayley--Hamilton theorem:
$ W_{38} = -W_5+W_6+1/2\, W_7 + W_8$.

According to the rules we have found in the strong sector, the
divergences of the octet weak sector for on--shell processes 
in the quenched case become:
\be
L^{8\, q}_{\mbox{\tiny{div}}} = W_4+\frac{1}{2}
W_6-\frac{3}{4} W_7  +\frac{1}{4}W_8 -\frac{1}{2} W_{11} \; \; .
\ee 
For the sake of clarity we list the definitions of the $W_i$ terms needed
here:
\ba
&&W_4=\langle\Delta u_\mu\rangle \langle u^\mu u_\nu u^\nu \rangle\, ,
~~~~~W_6= \langle\Delta u_\mu\rangle \langle\chi_+  u^\mu \rangle\, ,
~~~~~W_7=  \langle\Delta\chi_+\rangle  \langle u_\mu u^\mu \rangle\, ,
\nonumber\\
&&W_{8}=  \langle\Delta u_\mu u^\mu \rangle\langle\chi_+\rangle \, ,
~~~~~W_{11}=  \langle\Delta \chi_+\rangle \langle\chi_+\rangle\, .
\ea
We have verified that the contribution of the singlet sector to the
divergences indeed cancels the terms with a negative power of $N$. It
would remain to be checked by an explicit calculation 
that the inclusion of the fermionic 
ghosts removes the terms with the linear flavour
dependence \cite{weak}, as one expects.

Finally, we derive the contribution of quenched chiral logs in this
sector, according to the simple recipe given in the previous appendix.
The answer is very simple: there are no quenched chiral logarithms
in the on--shell octet weak sector at one loop. 
The reason is that if one expands the term proportional to $c_2$
in the lowest order Lagrangian around the background field, there
are no terms with $\xi_0^2$.
The situation would be different considering also off--shell
processes, since the expansion of the $c_5$ weak mass term generates the
singlet term $c_5
\xi_0^2 \langle \Delta \chi_+ \rangle$. The conclusion here is that
the quenched chiral logs can be reabsorbed in a redefinition of
$c_5$.

\renewcommand{\theequation}{\Alph{section}.\arabic{equation}}
\setcounter{equation}{0}
\section{One--loop functions}

\label{APP3}

Here we list the functions $J(q^2), I_1(q^2), I_2(q^2)$ which appear in 
the one--loop corrections to the quenched observables analyzed in this
paper. $ I_1(q^2)$ and $I_2(q^2)$ are not generated in 
standard CHPT and arise from the insertion of the $\alpha$ and $m_0^2$ 
vertices in any internal singlet line.

The one--loop function $J(q^2)$ in Minkowski space--time, 
is given by
\begin{equation}
J(q^2) = \frac{1}{i}\int \frac{d^dl}{(2\pi)^d}
\frac{1}{\left(M^2-l^2\right) 
\left(M^2-(l-q)^2 \right)} \; \; ,
\end{equation}
and
\be
{J}(q^2)=J(0)+\bar{J}(q^2) \; \; , 
\ee
where $J(0)$ contains the divergent part
\ba
J(0)&=&-2\lambda -{1\over 16\pi^2}\left ( \ln{M^2\over \mu^2} +1\right
) +O(d-4) \; \; ,
\nonumber\\ 
&&\nonumber\\
 \lambda &=& \frac{\mu^{d-4}}{16
  \pi^2}\left[\frac{1}{d-4}-\frac{1}{2}\left( \ln 4\pi
    +\Gamma^\prime(1)+1\right) \right] \; \; ,
\ea
and $\bar{J}(q^2)$ is finite. The explicit expression of $\bar{J}(q^2)$
for $d=4$ is:
\be
\bar{J}(q^2)= {1\over 16\pi^2}\left\{\sigma\ln{\sigma -1\over \sigma 
    +1}+2\right\} \; \; ,
\ee
where $\sigma =\sqrt{1-4M^2 / q^2}$.
The UV finite functions $I_1(q^2)$ and $I_2(q^2)$ are given by
\ba
I_1(q^2)&=&  \frac{1}{i}\int \frac{d^dl}{(2\pi)^d} \frac{1}{\left(M^2-l^2
\right)^2\left(M^2-(l-q)^2 \right)} \; \; ,    \nonumber\\
I_2(q^2)&=& \frac{1}{i}\int \frac{d^dl}{(2\pi)^d} \frac{1}{\left(M^2-l^2
\right)^2\left(M^2-(l-q)^2 \right)^2} \; \; .
\ea
They are the F.T. of the functions $\tilde{I}_1(z)$ and $\tilde{I}_2(z)$:
\ba
I_1(q^2) &=& \frac{1}{i} \int d^dz\, e^{iqz}\, \tilde{I}_1(z)\; \; ,
 \nonumber\\
I_2(q^2) &=&\frac{1}{i} \int d^dz\, e^{iqz}\, \tilde{I}_2(z) \; \; .
\ea
The explicit expression of $I_1(q^2)$ and $I_2(q^2)$ for $d=4$ is as follows
\ba
I_1(q^2) &=& {1\over 16\pi^2}{1\over q^2}{1\over \sigma}\ln
{ \sigma -1 \over  \sigma +1} \; \; , \nonumber\\
I_2(q^2) &=& - {1\over 8 \pi^2} {1\over q^4\sigma^2} \left[ 
1 + {q^2-2M^2 \over q^2 \sigma }
\ln { \sigma -1 \over  \sigma +1} \right] \; \; .
\ea
In the text we have used also these functions subtracted at $q^2=0$:
\be
I_{1,2}(q^2) = I_{1,2}(0) + \bar{I}_{1,2}(q^2) \; \;,
\ee
with 
\be
I_1(0) = {1 \over 32 \pi^2  M^2} \; \; , 
I_2(0) = {1 \over 96 \pi^2 M^4} \; \; .
\ee
When $q^2 > 4 M^2 $ both functions develop 
an imaginary part, which is given by 
\ba
\mbox{Im} ~I_1(q^2) &=& {1\over 16\pi}{1\over q^2}{1\over \sigma}\; \; ,
\nonumber\\ 
\mbox{Im} ~I_2(q^2) &=& -{1\over 16\pi}{2(q^2-2M^2)\over q^3\sigma^3} \; \; .
\ea
Notice that they diverge at $q^2= 4 M^2$.

\end{document}